\documentclass[a4paper,11pt]{article}
\usepackage[a4paper,top=2cm,bottom=2cm,left=3cm,right=3cm,marginparwidth=1.75cm]{geometry}
\usepackage[pdftex, pdftitle={Article}, pdfauthor={Author}]{hyperref}
\usepackage[utf8]{inputenc}
\usepackage{multicol}
\usepackage{float,array,multirow}
\usepackage{subcaption}
\usepackage{todonotes}
\usepackage{amsmath}
\usepackage[english]{babel}
\usepackage{xcolor}
\usepackage{pifont}

\usepackage[backend=biber,maxnames=2,style=numeric-comp,sorting=none]{biblatex}
\DeclareFieldFormat[article]{pages}{#1}
\DeclareBibliographyAlias{article}{std}
\DeclareBibliographyAlias{online}{std}
\DeclareBibliographyAlias{book}{std}
\DeclareBibliographyDriver{std}{
  \usebibmacro{bibindex}
  \usebibmacro{begentry}
  \usebibmacro{author/editor+others/translator+others}
  \setunit{\labelnamepunct}\newblock
  \newunit\newblock
  \usebibmacro{date}
  \newunit\newblock
  \usebibmacro{journal}
  \newunit\newblock
  \printfield{volume}
  \newunit\newblock
  \printfield{pages}
  \newunit\newblock
  \usebibmacro{finentry}}

\usepackage{lscape}
\usepackage{rotfloat}

\usepackage{hyperref}

\usepackage{graphicx}

\usepackage[percent]{overpic}

\begin{document}
\begin{center}
\section*{Hollow toroidal rotation profiles in strongly electron heated H-mode plasmas in the \mbox{ASDEX Upgrade} tokamak}
C. F. B. Zimmermann\textsuperscript{a b *}, R. M. McDermott\textsuperscript{a}, C. Angioni\textsuperscript{a}, B. P. Duval\textsuperscript{d}, R. Dux\textsuperscript{a},\\E. Fable\textsuperscript{a}, A. Salmi\textsuperscript{c}, T. Tala\textsuperscript{c}, G. Tardini\textsuperscript{a}, T. Pütterich\textsuperscript{a},\\and the ASDEX Upgrade\textsuperscript{e} team\\
\vspace{0.4cm}
\textit{\textsuperscript{a} Max Planck Institute for Plasma Physics, 85748 Garching, Germany}\\
\textit{\textsuperscript{b} Department of Applied Physics and Applied Mathematics, Columbia University, New York 10027, USA}\\
\textit{\textsuperscript{c} VTT, P.O. Box 1000, FI-02044 VTT, Finland}\\
\textit{\textsuperscript{d} Ecole Polytechnique Fédérale de Lausanne, SPC, CH-1015 Lausanne, Switzerland }\\
\textit{\textsuperscript{e} see the author list of H. Zohm et al., 2024 Nucl. Fusion, 10.1088/1741-4326/ad249d}\\
\vspace{0.4cm}
\textsuperscript{*} E-mail of the corresponding author: benedikt.zimmermann@ipp.mpg.de
\end{center}
\vspace{0.8cm}
\begin{abstract}

\noindent This work investigates toroidal momentum transport in type-I ELMy H-mode plasmas in the {ASDEX Upgrade} tokamak, focusing on the formation of hollow rotation profiles under strong electron cyclotron resonance heating (ECRH). Applying the established momentum transport analysis framework to a neutral beam injection (NBI) modulation experiment, momentum transport coefficients were inferred self-consistently. This was done for phases with dominant NBI heating and with additional strong ECRH, during which the rotation profile severely collapsed without significant changes in the externally applied torque. The experimental rotation profiles were accurately reproduced, confirming the robustness of the inferred diffusive, convective, and residual-stress contributions. While the Prandtl number and inward Coriolis pinch remained comparable between phases, the NBI+ECRH phase exhibited a strong counter-current intrinsic torque. Linear gyrokinetic simulations indicate a transition from ion-temperature-gradient (ITG) turbulence to an ITG–trapped-electron-mode (TEM) mixed regime under strong ECRH, consistent with the observed counter-current intrinsic torque and particle pinch behavior. Additional high-ECRH discharges with modified density demonstrated that hollow rotation profiles emerge from a balance between counter-current intrinsic torque and inward convective momentum transport, strongly influenced by the pedestal-top rotation level, which is dominantly set by variations in the pedestal-top density. These findings highlight the importance of intrinsic torque and inward convection for maintaining favorable rotation profiles in future low-torque tokamak scenarios and motivate further exploration of edge torque generation mechanisms.
\end{abstract}

\section{Introduction}

Present-day tokamaks, which typically rely on strong external torque sources provided by neutral beam injection (NBI) heating, often exhibit strongly peaked toroidal plasma rotation profiles. Plasma rotation has been shown to influence impurity transport \cite{Angioni_2012, casson2013, angioni2014tungsten, Casson_2015, Angioni2015}, stabilize turbulence, and enhance energy confinement \cite{biglari1990influence, Hahm1994, Waltz_1995_PoP, Hahm1995, Burrell1997, Terry2000}. In particular, rotation and its shear in the inner plasma core can suppress the locking and onset of magnetohydrodynamic (MHD) modes \cite{Strait1995, Garofalo2002, Reimerdes2007, Politzer2008, buttery2008, de_Vries_2011}.\par

A specific challenge arises in the presence of so-called \textit{hollow} rotation profiles, in which the radial gradient of the toroidal rotation changes sign within the plasma core, causing the rotation profile to roll over. These profiles are characterized by low core rotation values and correspondingly weak rotational shear, which can be prone to MHD instabilities \cite{Bardoczi_2024}. In the absence of already developed MHD modes, hollow rotation profiles are generally assumed to result from turbulent momentum transport in this radial region of the plasma.\par

Momentum transport in tokamak plasmas is inherently complex, as it comprises not only diffusive and convective momentum fluxes, but also a so-called \textit{residual stress} component \cite{Diamond2009, Camenen_PRL_2009, Camenen2011, Stoltzfus-Dueck_2012}. This residual stress can give rise to an \textit{intrinsic torque} that is capable of breaking plasma rotation, even in the presence of external momentum input \cite{Yoshida_2008_PRL, Yoshida_2009_PRL, Solomon2009, Solomon2011, Solomon_2010, Ida_1995, Rice_1998, Lee2003, Rice2004}. Among the various contributions to momentum transport, the residual stress represents the largest source of uncertainty in predicting core rotation profiles for future tokamak reactors, in which neutral beam injection is either not envisioned (e.g. the SPARC design) or not the dominant heating technique (e.g. ITER).

Given these challenges, significant effort has been devoted to advancing the theoretical understanding of turbulent momentum transport \cite{Garbet_2004, Peeters2007_PRL, Hahm_PoP_2007, Strintzi2008, Hahm_PoP_2008, Diamond2009, Peeters2011, Stoltzfus-Dueck_2012}. The corresponding experimental validation is still ongoing, however, substantial progress has been achieved in recent years, particularly with respect to measurements and analysis in the plasma core \cite{Tala_2009, Tala_2011, tala2016_EPS, Mantica_2010, Tardini2009, Yoshida_2012_intermachine, Ida_1998, Degrassie2009PPCF, Ida_2014, Rice_2016, rice_2022, Zimmermann_2024}.

A study by McDermott \textit{et al.} investigated the effects of electron cyclotron resonance heating (ECRH) on toroidal rotation in H-mode discharges of the ASDEX Upgrade (AUG) tokamak \cite{mcdermott2011effect, mcdermott2011core}. It was shown that the application of core ECRH can significantly modify the plasma rotation profiles despite the absence of a modification of the externally applied torque. The observed changes could not be explained by preferential fast-ion losses or purely diffusive momentum transport. Instead, they were attributed either to outward convective momentum transport or to a counter-current intrinsic torque. At the time of that study, however, the numerical and experimental tools required to disentangle the individual momentum transport components were not sufficiently developed, preventing a definitive identification of an ECRH-driven counter-current intrinsic torque. Nevertheless, the combination of an inward Coriolis pinch and counter-current intrinsic torque was considered the most probable. While recent theoretical and experimental studies \cite{Camenen2011, Angioni_PRL_2011, Zimmermann_2024} have explained the emergence of counter-current intrinsic torque in mixed turbulence regimes involving both ion-temperature-gradient (ITG) and trapped-electron-mode (TEM) turbulence, it remained unclear under which conditions this mechanism leads to hollow rotation profiles, particularly, as hollow profiles proved difficult to reproduce in subsequent AUG experiments.

The present study aims to clarify these open questions by employing the momentum transport analysis framework developed at AUG \cite{Zimmermann2022, Zimmermann_NF_Letter, Zimmermann_NF_Isotope}. The reader is referred to these publications for a detailed description of the methodology and modeling approach. In particular, the approach in Ref. \cite{Zimmermann_NF_Isotope} is very similar to the one taken in this work. The analysis is based on plasma experiments with controlled torque perturbations, for example, but in general not limited to, modulation of NBI. Fourier decomposition techniques applied to the measured toroidal rotation profiles yield amplitude, phase, and time-averaged steady-state profiles. These allow for the separation of diffusive, convective, and residual stress contributions based on the characteristic temporal response of each transport channel. Experimental ion temperature and rotation profiles are obtained using charge-exchange recombination spectroscopy (CXRS), under the assumption that the measured main impurity species boron is representative of the main ions \cite{Viezzer2012, McDermott2017}.

The transport analysis is performed using the ASTRA transport code \cite{ASTRA_Reference_Paper, Fable2013, Tardini2026ASTRA8}, which solves the toroidal momentum conservation equation \cite{fable2015toroidal} and predicts plasma rotation for a prescribed set of transport coefficients. Statistical optimization algorithms are employed to iteratively adjust these coefficients until agreement between the modeled and experimentally measured rotation profiles is achieved. Associated $1\sigma$ uncertainty estimates are assigned to the modeling solution, as discussed in \cite{Zimmermann_NF_Isotope}.

In the modeling, following an approach discussed in a very similar manner in Ref.~\cite{Zimmermann_NF_Isotope}, an experimental boundary condition is imposed at a normalized toroidal flux radius of approximately $\rho_\varphi = 0.8$. Variations in the precise radial location of this boundary condition have only a minor effect, provided that the analyzed profiles remain sufficiently smooth and that the boundary condition avoids regions affected by edge-localized modes (ELMs). Numerical experiments performed for the scenario discussed in this work, in which the boundary conditions were varied within the experimental uncertainties, show that the resulting effects are limited to the region close to the boundary and remain within the uncertainties of the inferred transport coefficients.

The flux-surface-averaged radial transport of toroidal angular momentum is expressed as
\begin{equation}
    \Gamma_\varphi = - m n R \left( \chi_\varphi\,\frac{\partial v_\varphi}{\partial r} - V_\mathrm{c}\, v_\varphi \right) + \Pi_\text{Rs} \, ,
    \label{eq:momentum_flux}
\end{equation}
where $m$ denotes the mass of the main ion species, $n$ its density, and $R$ the flux-surface-averaged local major radius. The coefficient $\chi_\varphi$ represents the momentum diffusivity. The toroidal velocity is defined as $v_\varphi = \langle R \Omega_\varphi \rangle$, where $\langle \cdot \rangle$ denotes a flux-surface average and $\Omega_\varphi$ is the toroidal component of the angular rotation frequency. For sufficiently weak poloidal plasma rotation, as is typically the case in the plasma core of the discharges considered here, $\Omega_\varphi$ can be assumed to be constant on a given flux surface \cite{lebschy2017measurement}. If poloidal rotation becomes non-negligible, the experimentally measured toroidal rotation must be corrected prior to comparison with the parallel velocity used in the modeling framework. The operator $\partial/\partial r$ denotes differentiation with respect to the minor radius $r$, $V_\mathrm{c}$ is the convective velocity, and $\Pi_\text{Rs}$ represents the residual stress contribution. This residual stress gives rise to an intrinsic torque defined as 
\begin{equation}
    \tau_\text{int} = - \partial V / \partial r \, \Pi_\text{Rs} \text{,}
\end{equation}
where $V$ is the plasma volume enclosed by a given flux surface.

Within the transport modeling framework, the momentum diffusivity $\chi_\varphi$ is linked to the experimentally inferred ion heat diffusivity $\chi_i$ through a linear dependence on the normalized toroidal flux coordinate $\rho_\varphi$. Their ratio defines the Prandtl number, $Pr = \chi_\varphi / \chi_i$, which is typically found to be of order unity. This observation supports the assumption of strong coupling between turbulent momentum and heat transport \cite{Peeters2005, waltz2007coupled, weiland2009symmetry, peeters2006toroidal, Holod_2008, Holod_2010, Strintzi2008, Peeters2011}. The ion heat diffusivity $\chi_i=-Q_i/(n\nabla T_i)$ (with the ion heat flux $Q_i$ and the ion temperature gradient $\nabla T_i$) is obtained from power balance calculations based on experimental measurements, ensuring that variations in turbulence intensity, such as those induced by modulation of NBI heating, are consistently included in the momentum transport modeling. The dimensionless quantity $-R V_\mathrm{c} / \chi_\varphi$ is commonly referred to as the pinch number. The radial profiles of the pinch number and residual stress are prescribed using quartic polynomials and are chosen to vanish at $\rho_\varphi = 0$ for continuity. Both convective and residual stress fluxes are chosen to scale with the inferred $\chi_\varphi$ to account properly for possible modulations in the turbulence amplitude, as discussed in detail in Refs. \cite{Zimmermann_NF_Letter,Zimmermann_NF_Isotope}. The externally applied torque and heat fluxes from NBI are calculated using the Monte Carlo code NUBEAM \cite{pankin2004}, which is embedded within the TRANSP framework \cite{TRANSP_Reference_Paper, hawryluk1981empirical} and incorporated self-consistently into the momentum balance equation. For modeling the heat fluxes of the ECRH, the TORBEAM code is included in the TRANSP workflow \cite{Poli2018}.

In the remainder of this paper, a plasma discharge with strong ECRH is introduced in Section \ref{sec:experiment}. Section \ref{sec:modeling} presents a thorough momentum transport analysis of this discharge. Section \ref{sec:emergence} discusses important ingredients for the formation of hollow rotation profiles by contrasting the presented experiments with additional data. The paper is concluded by a summary.

\section{Experimental Description}
\label{sec:experiment}

In this section, the AUG plasma discharge \#29216 is introduced. It is a standard type-I ELMy H-mode plasma in a toroidal field of $B_\varphi = -2.5$ T, a plasma current of $I_p = 0.6$ MA, a line-averaged core electron density of $7.2\cdot10^{19}$ m\textsuperscript{-3}, with a lower-single-null, favorable-drift configuration.

\begin{figure}
    \centering
    \includegraphics[width=0.8\textwidth]{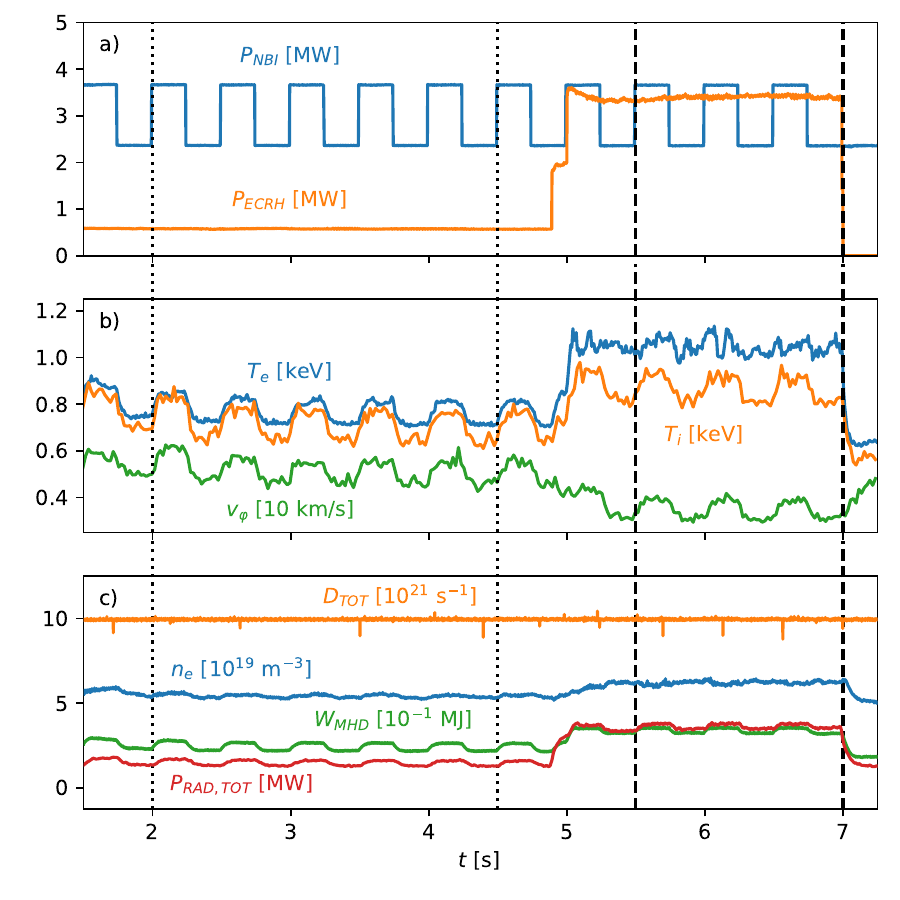}
    \caption{Most important time traces of \#29216. The studied discharge phases are indicated by vertical lines: the dominant NBI heating phase ($2.0-4.5$ s) by dotted lines and the phase with strong ECRH (5.5--7.0 s) by dashed lines. All time traces of quantities that can be radially resolved are sampled at mid-radius.}
    \label{fig:timetraces}
\end{figure}

Figure~\ref{fig:timetraces} shows time traces of the most relevant physics quantities. As illustrated by the blue NBI power trace in Fig.~\ref{fig:timetraces}(a), a single full-power steady beam ($2.4$ MW, $59$ kV) is used for plasma heating and serves as the diagnostic beam for CXRS \cite{Viezzer2012,McDermott2017}. In addition, a reduced-power beam at approximately $1.3$~MW ($65$ kV) is modulated at a frequency of $2$~Hz for the momentum transport analysis. The discharge consists of a phase with low ECRH power to avoid impurity accumulation \cite{Angioni_2017,Stober_2003}, $P_\mathrm{ECRH}=0.6$~MW (from $2.0$ to $4.5$~s, hereafter referred to as the ``NBI phase''). It is followed by a phase in which strong ECRH, $P_\mathrm{ECRH}=3.4$~MW, is applied (from $5.5$ to $7.0$~s, referred to as the ``NBI+ECRH phase''). The time intervals selected for analysis are indicated by dotted and dashed vertical lines, respectively. They encompass five modulation cycles for the NBI phase and three modulation cycles for the NBI+ECRH phase. The low ensemble does not pose an issue for the analysis due to the stable background quantities, as discussed in the following.

As shown in Panel~(b), the NBI modulation produces the desired modulation of the toroidal plasma rotation (green curve) and exhibits a significant collapse despite unchanged NBI settings and, therefore, constant externally applied torque. However, it also leads to an undesired modulation of the electron and ion temperatures, shown by the blue and orange curves, respectively. The increased total heating power during the second discharge phase results in higher temperatures, see the electron temperature $T_e$ in orange and the ion temperature $T_i$ in blue in Panel (b). Although ECRH directly heats only the electrons, collisional coupling, particularly at higher plasma densities, leads to a corresponding increase in ion temperature as well. The higher electron temperatures lead to enhanced radiated power $P_\mathrm{rad}$, see red curve in Panel~(b). The higher input power leads to an increase in the stored magnetic energy $W_\mathrm{MHD}$ (green curve). The plasma equilibrium shape was found to be stable during the discharge.

The gas fueling rate is held constant throughout the discharge at approximately $10 \times 10^{21}$~D atoms/s, see orange curve in Panel (c). The electron density, shown by the blue curve in the same Panel, remains relatively constant during the NBI phase but increases during the NBI+ECRH phase. Given the constant fueling rate, this density rise is attributed to changes in particle transport and suggests the presence of an inward particle pinch, as discussed later. Since the density remains approximately constant over the modulation itself, variations in the plasma angular momentum $L_\varphi = m_i n_i R v_\varphi$ are driven solely by changes in the toroidal rotation. This motivates performing the model-experiment optimization using the rotation velocity rather than the angular momentum itself, even though the latter is the physically conserved quantity. The precise temporal evolution of the density is included in the modeling. The effective charge number $Z_\mathrm{eff}$ is reconstructed from bremsstrahlung measurements and remains approximately constant at $Z_\mathrm{eff} \simeq 1.3$ during both discharge phases, not shown for brevity.

The sawtooth (ST) inversion radius increases from $\rho_\varphi = 0.18$ in the NBI phase to $\rho_\varphi = 0.22$ in the NBI+ECRH phase, confirmed by manual analysis of electron cyclotron emission radiometry (ECE) data. This change is accounted for later in the analysis by a slight adjustment of the analysis window, avoiding the regions found to be affected by the ST, and is attributed to the modified core heating and possible alterations of the current profile induced by ECRH. The sawtooth frequency exceeds $50$~Hz in both phases, and if they were to influence the core profiles, the analysis would average over their effects. Aside from sawtooth activity, no strong core MHD modes are identified in the analysis of the magnetic probe signals. ELMs occur at high frequency and, similar to sawteeth, are not temporally resolved by the CXRS diagnostics, which operate at a time resolution of $10$~ms. Consequently, the transport analysis excludes the corresponding radial regions in order to focus on turbulent transport effects. Owing to the low NBI modulation frequency of $2$~Hz, neither sawteeth nor ELMs are expected to introduce spurious higher harmonics into the modulation response.

\begin{figure}[t]
    \centering
    \includegraphics[width=0.9\textwidth]{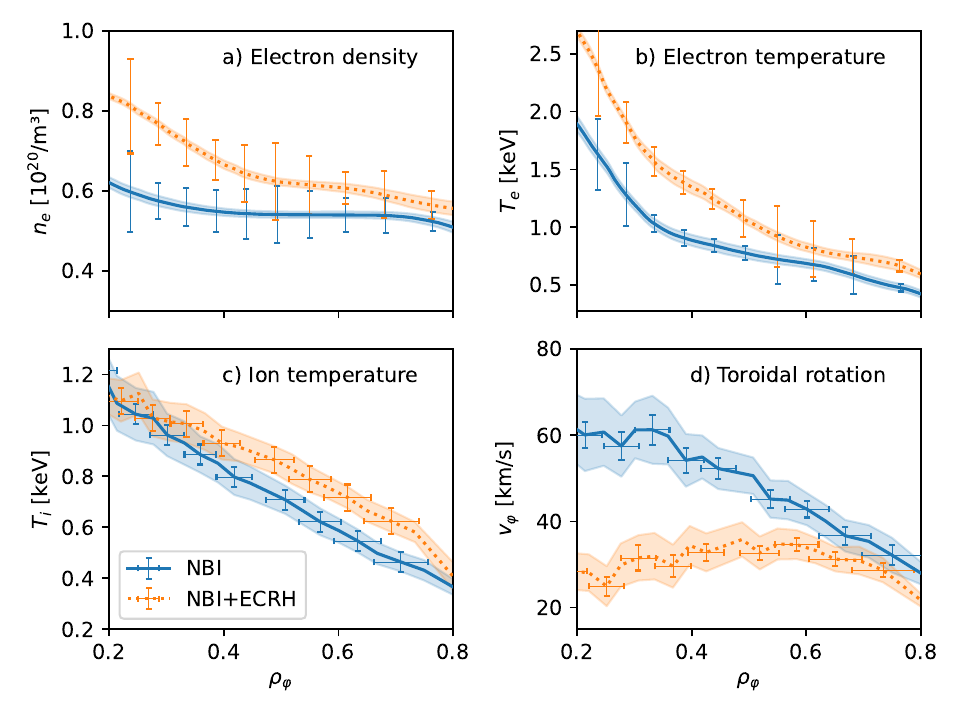}
    \caption{Kinetic profiles for \#29216 for the discharge phase with dominant NBI heating (2.0--4.5s, blue solid) and for the phase with NBI+ECRH (5.5--7.0s, orange dotted). The error bars represent the statistical uncertainty in data reconstruction, while the band structures indicate the standard deviation across the analyzed time windows for each radial point.}
    \label{fig:kinetic_profiles}
\end{figure}

Beyond the induced 2 Hz perturbation, all the average values of discussed quantities remain stable over the analyzed discharge phases. This motivates the analysis and comparison of time-averaged kinetic profiles. They are shown in Fig. \ref{fig:kinetic_profiles} with the electron density in Panel (a) and the electron temperature in Panel (b). The electron density and temperature profiles are reconstructed within the Integrated Data Analysis framework \cite{IDA_paper}, which integrates measurements from lithium beam emission spectroscopy \cite{LIB_paper_1}, laser interferometry \cite{mlynek2010design}, ECE, and Thomson scattering diagnostics \cite{Thomson_System_AUG}. The error bars represent the statistical uncertainties arising from the Bayesian inference used in the reconstruction. A detailed discussion of the treatment of uncertainties in this analysis methodology is provided in \cite{Fischer_2020}.

The electron density is found to increase significantly during the NBI+ECRH phase, indicating that the additional ECRH drives the plasma into an ITG-TEM mixed-mode turbulence regime characterized by an inward particle pinch, resulting from modifications to the microturbulence \cite{Angioni_2005,Fable_2008,Fable_2010, angioni2011gyrokinetic,Angioni_PRL_2011}. This observation suggests that the applied level of ECRH power, in combination with the still substantial ion heating provided by NBI and the relatively high plasma density, is insufficient to establish a deeply TEM-dominated regime during the second discharge phase.

A rise in the electron temperature is observed following the increase in ECRH power, and a corresponding increase in the ion temperature is shown in Panel~(c), resulting from collisional energy exchange between electrons and ions. In contrast, the toroidal rotation profile displayed in Panel~(d) exhibits a pronounced flattening inside mid-radius. This leads to a hollow rotation profile characterized by a reversal of the radial gradient and a near-vanishing rotational shear throughout the core, despite the application of substantial NBI torque. Such behavior indicates a strong change of momentum transport properties under ECRH in a mixed-mode turbulence regime.

The error bars shown for $T_i$ and $v_\varphi$ reflect the statistical uncertainties associated with the fitting of CXRS measurements. For the toroidal rotation velocity, additional systematic uncertainties arising from calibration are included. In contrast, systematic uncertainties in $T_i$ are negligible, as the spectrometer dispersion relation is accurately characterized. Finally, for all four quantities shown, the shaded bands represent the standard deviation over the analyzed discharge phase, providing an indication of the stability of the signals under modulation.

\begin{figure}[t]
    \centering
    \includegraphics[width=0.9\textwidth]{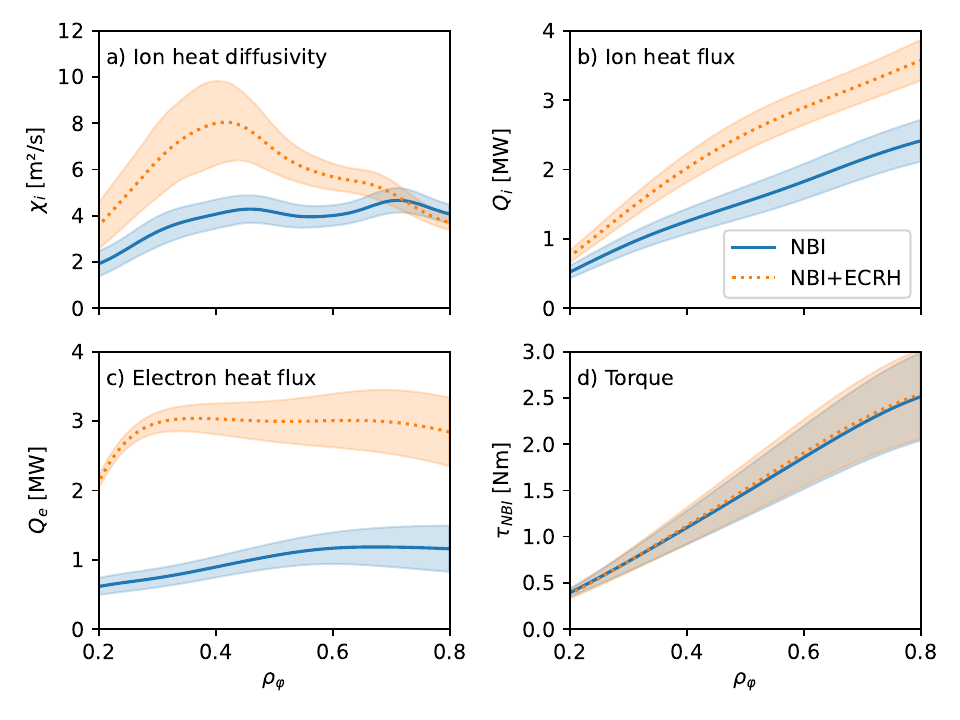}
    \caption{Ion heat diffusivity (a) from ASTRA calculations. Ion heat fluxes (b), electron heat fluxes (c), and applied torque (d) from TRANSP calculations, shown for \#29216 for the discharge phase with dominant NBI heating ($2.0-4.5$ s, blue solid) and for the phase with NBI+ECRH ($5.5-7.0$ s, orange dotted). The band structures indicate the standard deviation across the analyzed time windows for each radial point, mainly from the NBI modulation.}
    \label{fig:fluxes}
\end{figure}

To enable a more detailed interpretation of the heat and torque fluxes, Fig.~\ref{fig:fluxes} presents the relevant quantities. Panel~(a) shows the ion heat diffusivity $\chi_i$ obtained from power-balance analysis. As illustrated in Panel~(b), the increased ion heat flux during the NBI+ECRH phase contributes to higher $\chi_i$ in the inner core, while in the outer core the combination of modified density, ion temperature gradient, and heat flux leads to similar values of $\chi_i$. The increased ion heat flux is primarily driven by an enhanced collisional heat exchange (not shown for brevity). Panel~(c) highlights the pronounced effect of ECRH on the electron heat flux $Q_e$, fueling the collisional heat exchange. In contrast, the externally applied torque from NBI remains nearly unchanged between the two phases, as shown in Panel~(d). As the density increases in the NBI+ECRH phase, it is assumed that the overall rotation level decreases when a similar torque acts on a larger number of particles. This can be seen as a shift in the rotation values at $\rho_\varphi = 0.8$ in Fig. \ref{fig:kinetic_profiles}(d). The variation in the inner core density, however, is around $20$\%, while the rotation differs nearly by a factor of $2.5$. This suggests that the deviations in the core rotation are a result of additional changes in the effective momentum transport.


\begin{figure}
    \centering
    \includegraphics[width=0.85\textwidth]{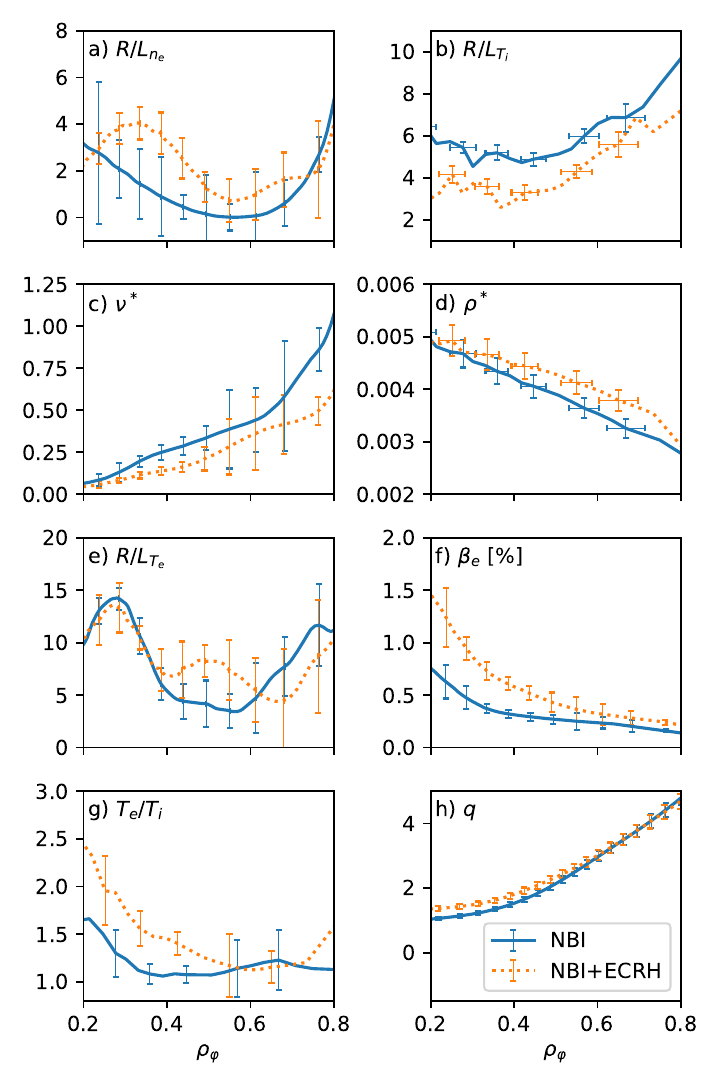}
    \caption{Comparison of dimensionless parameters in the plasma core of \#29216 for the discharge phase with dominant NBI heating ($2.0-4.5$ s, blue solid) and for the phase with NBI+ECRH ($5.5-7.0$ s, orange dotted). The additional ECRH modifies multiple dimensionless parameters, which are assumed to modify turbulence and the resulting transport, in particular the logarithmic electron density and temperature gradients and the electron to ion temperature ratio.}
    \label{fig:core_gradients}
\end{figure}

A set of dimensionless parameters, assumed to govern transport physics \cite{Petty1998}, is shown in Fig.~\ref{fig:core_gradients}. The error bars shown for the gradient quantities are obtained from Gaussian process regression, with the Gaussian uncertainties approximated by the statistical uncertainties of the input profiles. For the remaining dimensionless parameters, the error bars are determined using analytical uncertainty propagation.

Panel~(a) shows the logarithmic electron density gradient, $R/L_{n_e} = -R \nabla n_e / n_e$. The steepening of the core density during the NBI+ECRH phase results in a correspondingly higher density gradient. The logarithmic ion temperature gradient, shown in Panel~(b), is uniformly shifted due to the overall higher $T_i$ values in the NBI+ECRH phase. However, no explicit modification of the shape of the profile is observed.

The normalized plasma collisionality, $\nu^* \sim n_e / T_e^2$, is shown in Panel~(c). Within error bars, $\nu^*$ remains similar for both discharge phases up to $\rho_\varphi = 0.7$, with the observed differences arising from combined changes in $n_e$ and $T_e$. Panel~(d) presents the normalized ion gyro-radius, $\rho^* = \rho_i / a = \sqrt{m_i T_i} / (e B a)$, where $a$ is the minor radius, $\rho_i$ the ion Larmor radius, and $e$ the elementary charge. As expected, $\rho^*$ increases in the NBI+ECRH phase due to the higher ion temperatures.

The logarithmic electron temperature gradient is shown in Panel~(e), exhibiting higher values around mid-radius during the NBI+ECRH phase. This behavior is a direct consequence of the increased electron heat flux. The electron plasma beta, $\beta_e \sim n_e T_e / B^2$, shown in Panel~(f), increases in the NBI+ECRH phase as a result of both density peaking and higher electron temperatures. Pronounced differences are also observed in the ratio of electron to ion temperature, $T_e/T_i$, shown in Panel~(g), which can be considered as a proxy for changes in the underlying turbulence regime. Panel~(h) displays the safety factor profile, which, in cylindrical approximation, can be calculated as $q = r/R \, B_\varphi / B_\theta$ and remains very similar between the two phases.

In summary, this initial comparison demonstrates that the analyzed plasma discharge is stable and well suited for a momentum transport analysis. The NBI+ECRH phase exhibits several indicators consistent with the presence of an ITG-TEM mixed-mode turbulence regime, most notably an enhanced inward particle pinch. During this phase, the toroidal rotation profile becomes strongly hollow. While this behavior is partly consistent with the increased plasma inertia at constant externally applied torque, the comparisons presented above also point to changes in the underlying momentum transport dynamics. Several dimensionless core parameters that are known to influence transport are found to be modified by the additional ECRH. In the following section, a detailed momentum transport analysis is performed to quantify the impact of these changes on the transport coefficients.

\section{Momentum Transport Modeling}
\label{sec:modeling}

The momentum transport analysis framework introduced in previous studies \cite{Zimmermann_NF_Letter, Zimmermann_NF_Isotope} has been applied to both discharge phases. In both cases, the analysis yields an excellent reproduction of the experimentally measured profiles.

\begin{figure}
    \centering
    \includegraphics[width=0.9\textwidth]{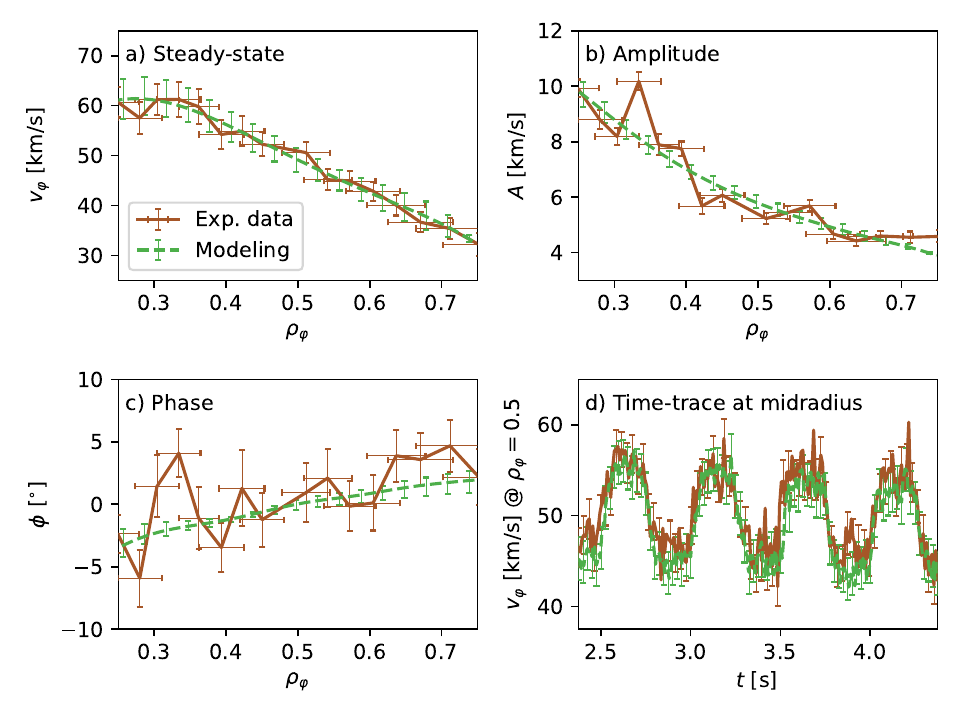}
    \caption{Modeling of the discharge phase with dominant NBI heating (\#29216, $2.0-4.5$ s).}
    \label{fig:fitting_NBI}
\end{figure}

\begin{figure}
    \centering
    \includegraphics[width=0.9\textwidth]{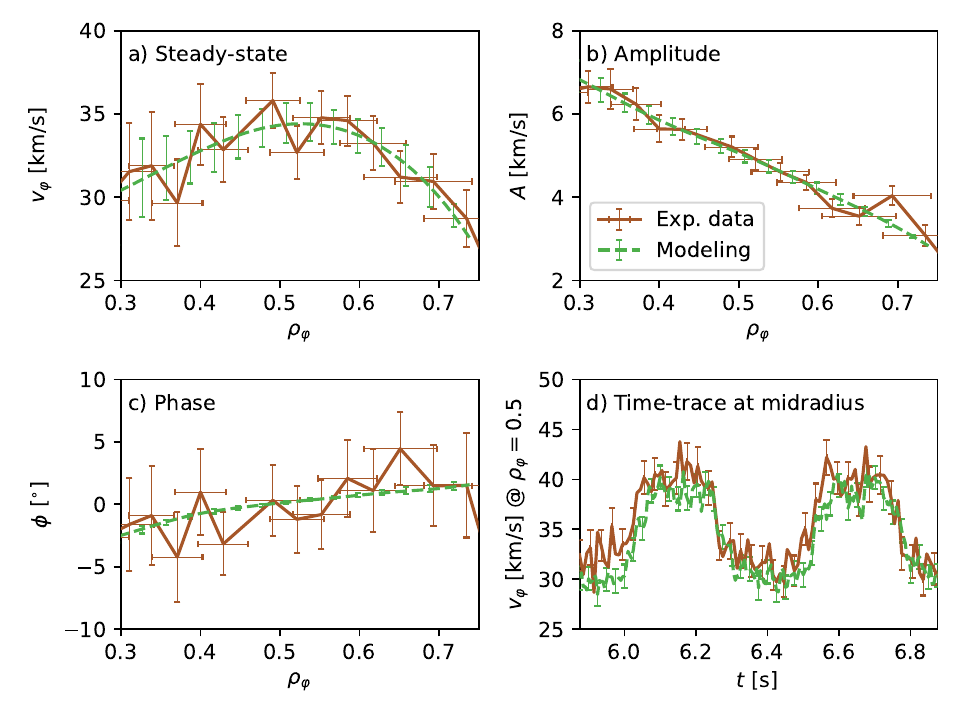}
    \caption{Modeling of the discharge phase with NBI+ECRH (\#29216, $5.5-7.0$ s).}
    \label{fig:fitting_ECRH}
\end{figure}

Figure~\ref{fig:fitting_NBI} shows the results for the first discharge phase with dominant NBI heating. The modeling results (shown in dashed green) reproduce the experimentally measured (brown solid) steady-state, see Panel~(a), amplitude, see Panel~(b), and phase profiles, see Panel~(c), with high accuracy. As demonstrated by the modeled time trace at mid-radius, Panel~(d), no spurious drift is observed, confirming that the inferred transport coefficients yield a consistent steady-state solution. The modeled rotation time trace exhibits a noticeable level of noise, which arises from the absence of temporal smoothing of the experimental kinetic profiles used as input to the modeling. This noise propagates into the rotation modeling through the coupling of the momentum diffusivity to the ion heat diffusivity, which is calculated based on these unsmoothed data. Figure~\ref{fig:fitting_ECRH} shows a very similar level of agreement for the discharge phase with NBI+ECRH, with both the radial profiles and the time traces being well reproduced. Consistent with the observed change in the sawtooth inversion radius, the fitting domain was slightly adjusted, being restricted to $\rho_\varphi = 0.25$ for the NBI phase and to $\rho_\varphi = 0.30$ for the NBI+ECRH phase.

A comparison of the amplitude profiles reveals systematically lower values during the NBI+ECRH phase, suggesting either enhanced diffusive transport or reduced convective contributions. This interpretation is further supported by the comparatively flatter phase profile observed in the NBI+ECRH case, indicative of increased diffusion in the NBI+ECRH phase \cite{Zimmermann2022}. The phase shift of the NBI-induced perturbation across the plasma core is small compared to both the error bars and local variations in the profile. This is a direct consequence of the relatively low modulation frequency of $2$~Hz and could be improved in future experiments by choosing higher modulation frequencies. The small phase shift observed here provides weaker constraints on the fitting. In contrast, previous work in Ref.~\cite{Zimmermann_NF_Isotope}, which used a $5$~Hz modulation, achieved steeper phase profiles, better constraining the modeling of $Pr$ and correspondingly yielding smaller uncertainties in the fitted transport quantities. As discussed later, this results in comparatively large uncertainties in the inferred Prandtl numbers, which subsequently propagate via the coupling of convective and residual stress terms to $\chi_\varphi$ into all extracted momentum transport coefficients.

\begin{figure}
    \centering
    \includegraphics[width=0.9\textwidth]{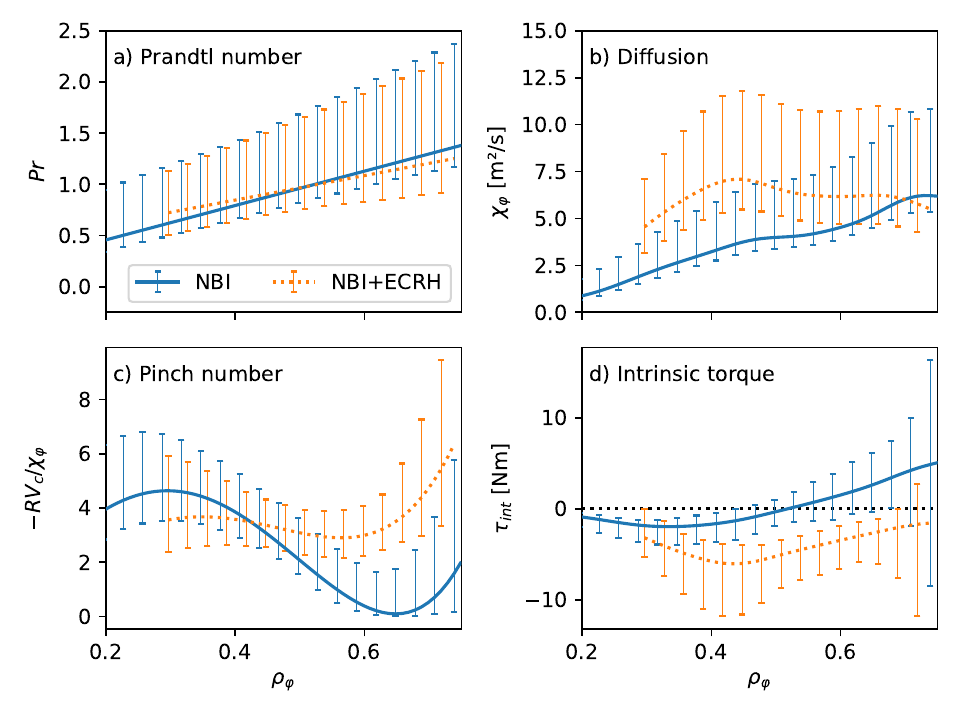}
    \caption{Comparison of the inferred transport coefficients for \#29216 for the discharge phase with dominant NBI heating ($2.0-4.5$ s, blue solid) and for the NBI+ECRH phase ($5.5-7.0$ s, orange dotted). Strong counter-current intrinsic torque is found in the NBI+ECRH phase.}
    \label{fig:transport_coefficients}
\end{figure}

Figure~\ref{fig:transport_coefficients} shows the inferred transport coefficients, with solid blue lines corresponding to the dominant NBI phase and dashed orange lines to the NBI+ECRH phase. The shown error bars correspond to a statistical $1\sigma$ confidence interval around the best-fit solution, reflecting the spread of the inferred quantities due to measurement and modeling uncertainties. A more detailed discussion of the underlying methodology is provided in Ref.~\cite{Zimmermann_NF_Isotope}.

Panel~(a) presents the Prandtl number profiles, which are very similar in both phases. This behavior is consistent with previous studies, which have shown that, in the absence of strong TEM-dominated turbulence, the Prandtl number depends only weakly on plasma parameters \cite{Strintzi2008, Kluy_2009, Zimmermann_2024}. In both cases, the Prandtl number increases with radius, in agreement with earlier work \cite{Zimmermann_NF_Letter, Zimmermann_NF_Isotope, Zimmermann_2024}, which suggests that the most significant scaling of the Prandtl number is with the inverse aspect ratio as a proxy for the trapped particle fraction, as discussed in more detail in Ref. \cite{Zimmermann_2024}. The asymmetric error bars arise from the very flat phase profiles: the modeled phase profiles can be flattened substantially before a meaningful increase in the fitting cost function occurs, whereas any additional steepening (corresponding to lower diffusion) quickly degrades the fit quality. If highly localized variations of the Prandtl number were present, they would likely be smoothed out by the linear radial model employed here. The use of higher-order polynomial representations is deferred to future work.

The absence of strong variation of the Prandtl number, together with the increased ion heat diffusivity observed in the NBI+ECRH phase (see Fig.~\ref{fig:fluxes}a), leads to a corresponding increase in the momentum diffusivity, see Panel (b). This finding is consistent with the flatter phase profiles, the reduced amplitude profiles, and the general expectation of enhanced turbulence levels due to the higher total power input during the NBI+ECRH phase.

As noted above, the relatively large uncertainties propagate through the analysis, resulting in sizeable error bars for the inferred pinch number profiles, shown in Panel~(c). In the NBI phase, the pinch number is largest in the inner core and decreases beyond mid-radius. This behavior closely resembles the profile of the logarithmic density gradient shown in Fig.~\ref{fig:core_gradients}(a) and is consistent with the theoretical concept of the Coriolis momentum pinch, which is found in theoretical \cite{Peeters2007_PRL} and experimental studies \cite{Tala_2009, Tala_2011, Zimmermann_2024} to scale most strongly with the logarithmic density gradient. A similar trend is observed in the NBI+ECRH phase, which exhibits slightly higher $R/L_{n_e}$ values in the outer core, reflected in a correspondingly larger pinch number. It should be noted that while the origin of the inward particle pinch emerging during the transition is, first of all, independent of the momentum convection, the modified density gradients obviously affect the Coriolis momentum pinch. In addition, the presented methodology does not disentangle advected and convected momentum. While the observed trends are in line with expectations, the relatively large error bars, together with the possible entanglement of advection and convection, warrant caution in the quantitative interpretation of the obtained results for the pinch number.

Finally, the assessed intrinsic torque values are shown in Panel (d). The discharge phase with dominant NBI shows slightly counter-current intrinsic torque values in the inner core, which become strongly co-current towards the edge of the analysis domain. In shape and size, they resemble previous analysis results of standard beam-heated H-mode discharges at DIII-D \cite{Solomon2009} and AUG \cite{Zimmermann2022,Zimmermann_NF_Isotope,Zimmermann_2024}. The corresponding curve for the NBI+ECRH phase shows more pronounced, stronger counter-current intrinsic torque, outside of error bars around mid-radius. While the profile shape suggests a sign reversal of the intrinsic torque back towards co-current towards the edge, the size of the error bars towards the fitting boundary does not enable to draw meaningful conclusions from these analysis results.

The observed modification of the intrinsic torque during the NBI+ECRH phase is consistent with the theoretical concept of residual stress generation arising from profile shearing effects and turbulent eddy tilting during the transition from ITG- to TEM-dominated turbulence \cite{Camenen2011}. In particular, the substantial variations in $R/L_{n_e}$, see Fig.~\ref{fig:core_gradients}(a), and in $R/L_{T_e}$, see Panel~(e), have been identified in previous studies \cite{Angioni_PRL_2011, Grierson_PRL_2017, Hornsby2018, Zimmermann_2024} as key drivers of counter-current intrinsic torque generation. While momentum transport and rotation physics differ fundamentally in stellarators due to the absence of axisymmetry (see discussion in Refs.~\cite[pp.~289]{hinton1976theory} and \cite[pp.~249]{helander2002collisional}), similarities can be identified with observations in the LHD device \cite{Ida_2021_LHD}. In LHD, TEM-dominated scenarios, associated with relatively flat density profiles, exhibit co-current rotation, whereas ITG-dominated scenarios, characterized by density profile peaking in the outer core, tend to exhibit counter-current rotation. This behavior is consistent with the correlation between intrinsic torque and kinetic profile gradients discussed in the present work and in the referenced studies. While this study is focused on H-modes, as L-modes cannot be investigated with the beam modulation technique due to the number of required beams, previous studies on these effects on AUG encompassed both L- and H-modes \cite{Angioni_PRL_2011,McDermott_2014,mcdermott2011effect} and showed that both confinement regimes exhibit the same physical mechanisms of intrinsic rotation generation, hollowing, and sign reversal through the relevant density gradients. In addition, this could also relate to previous work on LHD \cite{Ida2013ReversalTorque}, showing changes in the intrinsic torque during the formation of internal transport barriers, which, again, could be correlated with changes in the density gradients.

As in these experiments no scans in the ECRH power were conducted, and as the counter-current intrinsic torque is self-consistently set by the turbulence, mediated by the kinetic profiles, no global threshold for the applied ECRH power is expected to exist for the hollowing. It should be noted that, from a theoretical standpoint \cite{Camenen2011,Hornsby2018}, intrinsic torque generation is expected to correlate with second-order derivatives of the kinetic profiles, which are, however, difficult to measure accurately in the experiment. Therefore, first-order derivatives are usually used as a proxy in such an analysis. A definitive interpretation of these observations would require non-linear, global gyrokinetic simulations \cite{Hornsby2018, Wang_PRL_2009, Wang_PoP_2010, Grierson_PRL_2017}, which are beyond the scope of the present work. It should be noted that other potential sources of intrinsic torque, such as fast-ion losses, neutral drag, preferential ion losses \cite{Stoltzfus-Dueck_PRL_2012}, or neoclassical toroidal viscosity \cite{Zhu_2006,Garofalo_2008,COLE_2011,Shaing_NF_2015}, are expected to be negligible in the plasma core for the discharges considered here.

\begin{figure}
    \centering
    \includegraphics[width=0.9\linewidth]{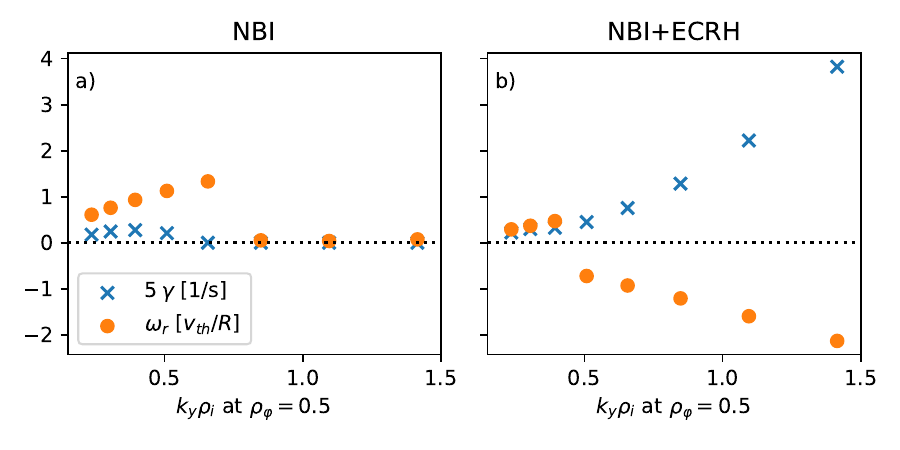}
    \caption{Turbulence growth rate $\gamma$ and real mode frequency $\omega_r$ scanned as a function of $k_y \rho_i$ using experimental input at mid-radius for both discharge phases of \#29216 for the discharge phase with dominant NBI heating ($2.0-4.5$ s, l.h.s.) and for the phase with NBI+ECRH ($5.5-7.0$ s, r.h.s).}
    \label{fig:GKW_spectra}
\end{figure}

To gain further insight into the underlying turbulence regime, the flux-tube version of the gyrokinetic GKW code \cite{Peeters2004, Peeters2009} was employed using experimental input profiles and an explicit time-integration (initial-value) approach. This allows the identification of the micro-instability with the highest linear growth rate $\gamma$, its dominant mode frequency $\omega_r$, and enables an analysis of the mode spectra. These calculations were done at mid-radius for both discharge phases, where the impact of the counter-current intrinsic torque appears to be strongest. The resulting mode spectra are shown in Fig.~\ref{fig:GKW_spectra} for the NBI phase in Panel~(a) and for the NBI+ECRH phase in Panel~(b). In the NBI phase, the mode frequency, see orange points, remains positive across the entire spectrum, indicating turbulence dominated by ITG modes. The growth rates, see blue crosses, are non-zero for smaller $k_y\,\rho_i$, remain finite over low to intermediate wavenumbers, and decrease to zero towards higher $k_y\,\rho_i$ (i.e. shorter wavelengths). In contrast, during the NBI+ECRH phase, the spectrum transitions from ITG to TEM (see the sign change in $\omega_r$), suggesting the presence of an ITG-TEM mixed-mode regime. This transition is accompanied by reduced $\omega_r$ at low $k_y\,\rho_i$ and increasingly negative values at higher $k_y\,\rho_i$. The growth rates extend to higher $k_y\,\rho_i$ and diverge approximately as $\gamma \sim k_y^2$. However, this increase at high $k_y$ is a known feature of such linear calculations in this spectral domain, with higher $k_y\,\rho_s$ contributing less to the transport. In direct comparison, the growth rates at low $k_y\rho_i$ are comparable in both phases, while they become larger in the NBI+ECRH phase at higher $k_y\rho_i$. The absolute values of $\omega_r$ are larger in the NBI phase at low $k_y\rho_i$, whereas at higher $k_y\rho_i$ they are of similar magnitude and eventually larger in the NBI+ECRH phase.

In summary, and together with the experimental observation of an enhanced inward particle pinch during the NBI+ECRH phase, this provides strong support for the assumption of mixed-mode turbulence, as established by previous works \cite{Angioni_2005,Fable_2008,Fable_2010,Angioni_PRL_2011}. Unfortunately, no turbulence measurements were available during these discharges to further support this modeling. Further gyrokinetic or gyrofluid modeling is left for future work. However, it should be noted that identifying the fastest growing linear modes (e.g. ITG/TEM) remains a meaningful and widely used approach to determine the dominant underlying turbulence regime. In the present case, the experimentally observed core rotation levels are sufficiently small that their impact on the equilibrium $E \times B$ shear, and consequently on mode characteristics, is limited. This supports the consistency of using linear GKW results to identify the instability responsible for the observed counter-current torque.

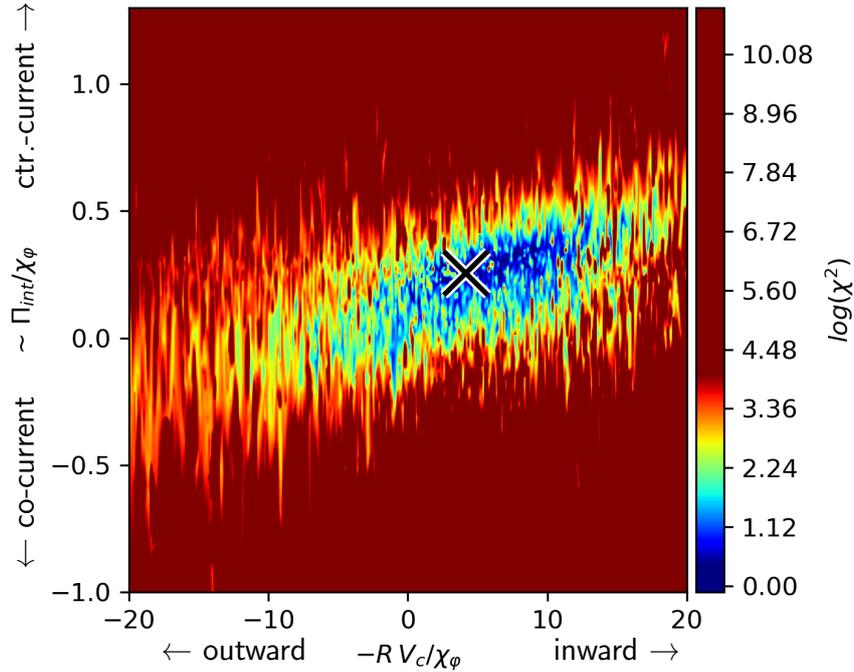
\begin{figure}[t]
    \centering
\begin{overpic}[width=0.8\linewidth]{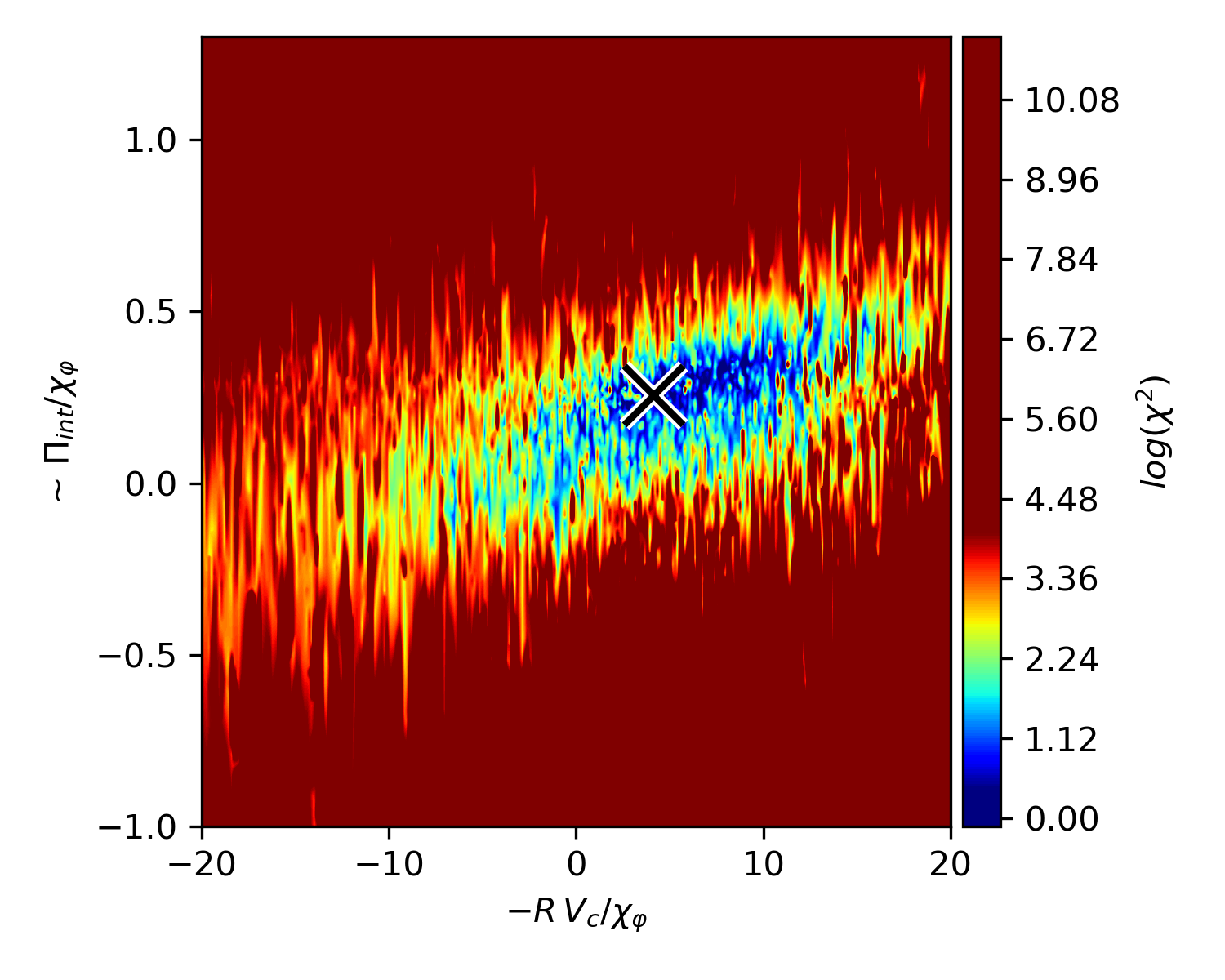}
    \put(63,5){\large\sffamily inward →}
    \put(20,5){\large\sffamily ← outward}
    
    \put(4,15){\rotatebox{90}{\large\sffamily ← co-current}}
    \put(4,57){\rotatebox{90}{\large\sffamily ctr.-current →}}
\end{overpic}
    \caption{Cost function values of the modeling optimization for the NBI+ECRH phase (\#29216, $5.5-7.0$~s). The tested pinch number (at mid-radius) is shown on the x-axis and the scaling parameter of the residual stress flux on the y-axis (at mid-radius). The final model solution is indicated by a black marker. The ``valley'' of acceptable solutions extends predominantly toward positive pinch numbers, suggesting inward convective momentum transport and peaking of the rotation profiles. Positive residual stress values yield a counter-current intrinsic torque contribution.}
    \label{fig:errorspace}
\end{figure}

While previous work by McDermott \textit{et al.} could not rule out the possibility of hollow rotation profiles arising from outward convective momentum transport, the analysis of the NBI+ECRH phase was repeated here without imposing any constraints or boundaries on the convective transport channel. In this analysis, all solutions featuring outward convection and co-current intrinsic torque were found to be inferior in terms of their cost-function values compared to solutions characterized by inward convection combined with counter-current intrinsic torque.

This result is illustrated in Fig.~\ref{fig:errorspace}, which shows the scanned $\chi^2_\text{red}$ cost-function values as a function of the pinch number (x-axis) and the fitting parameter of the residual stress flux (y-axis). A pronounced valley-like structure is observed, extending predominantly toward positive (co-current, inward) pinch number values, with the best-fit solution indicated by a black marker. Despite the noticeable noise in the cost-function landscape, arising from the simultaneous variation of diffusive transport parameters during the scan, this result, together with the final optimized solution, provides strong experimental evidence that a scenario involving inward convection and counter-current intrinsic torque offers a superior description of the experimental observations.

\begin{figure}
    \centering
    \begin{overpic}[width=0.9\linewidth]{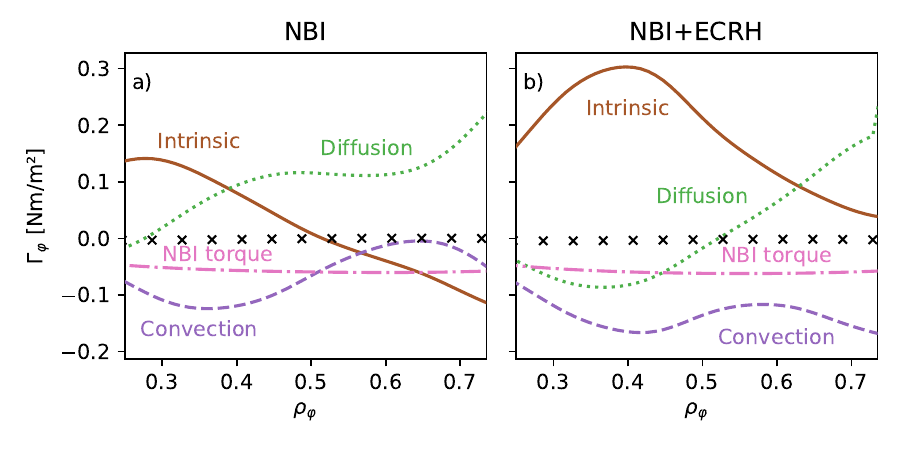}
        \put(99,9){\rotatebox{90}{\normalsize\sffamily ← inward}}
        \put(99,32){\rotatebox{90}{\normalsize\sffamily outward →}}
    \end{overpic}
    \caption{Comparison of the momentum fluxes for both discharge phases for \#29216 for the discharge phase with dominant NBI heating ($2.0-4.5$ s, l.h.s.) and for the phase with NBI+ECRH ($5.5-7.0$ s, r.h.s.). The black markers indicate the sum of all momentum flux contributions. Values close to zero demonstrate the temporal stability and self-consistency of the modeled solutions. Positive flux values correspond to an outward flux, which yields an apparent counter-current torque.}
    \label{fig:flux_comparison}
\end{figure}

Figure \ref{fig:flux_comparison} compares the fitted momentum flux contributions for both discharge phases using the same units, allowing their relative importance to be directly assessed. Positive values correspond to counter-current, outward momentum flux. Panel~(a) shows the values for the NBI phase, in which strong intrinsic and diffusive fluxes are present, while the externally applied momentum flux from NBI plays only a minor role in the inner core.

Panel~(b) presents the corresponding results for the NBI+ECRH phase on the same y-axis scale. In the NBI+ECRH phase, the counter-current intrinsic torque clearly dominates the overall momentum balance. Notably, the sign reversal of the rotation gradient leads to a situation in which the diffusive momentum flux acts to ``fill'' the hollow rotation dip, resulting in an apparent co-current diffusive flux. The convective contribution is also enhanced due to the larger pinch number and acts to partially counterbalance the collapse of the rotation profile in the NBI+ECRH phase. If, for example, the same transport coefficients inferred for the NBI+ECRH phase were applied to a plasma with higher overall rotation boundary values, the convective fluxes could be significantly amplified, as the convective momentum flux increases with the background plasma rotation. In such a scenario, these enhanced convective fluxes could balance the strong counter-current intrinsic torque and thereby mitigate, or even prevent, the formation of a hollow rotation profile.

In summary, the modeling reproduces the experimentally measured rotation profiles with high accuracy. The diffusive and convective momentum transport contributions respond as expected to the changes in the plasma scenario. The intrinsic torque reacts strongly to the modified gradient profile shapes and turbulence characteristics induced by the additional ECRH, which drives the plasma into an ITG-TEM mixed-mode turbulence regime, as confirmed by dedicated gyrokinetic calculations. This results in strong counter-current intrinsic torque and the formation of hollow rotation profiles. The convective fluxes are found to provide the best agreement with the experimental data when directed inward and could become significant in the NBI+ECRH phase, particularly if the pedestal-top rotation boundary values were higher.

\section{Formation of Hollow Rotation Profiles}
\label{sec:emergence}

While recent work based on the same analysis framework has already identified counter-current intrinsic torque in an ITG-TEM mixed-mode turbulence regime, it remained unclear under which conditions a sign change in the rotation gradient occurs and when distinctly hollow rotation profiles are formed \cite{Zimmermann_2024}. To address this question, a small set of discharges was designed to closely reproduce discharge \#29216 while allowing for controlled parametric variations. This effort resulted in discharges \#42339 and \#42340, which closely match \#29216 in terms of applied heating power, plasma current, magnetic field, equilibrium, and pedestal parameters.

For these discharges, it was not possible to perform a full torque perturbation analysis, possibly because the modulation beam power was reduced too strongly ($0.7$ MW, $55$ kV) and higher modulation frequencies were chosen, both leading to insufficient penetration into the plasma core and insufficient modulation amplitude profiles for analysis. Nevertheless, the time-averaged steady-state profiles provide valuable additional insight into the conditions under which hollow rotation profiles emerge. Future work could improve the modulation scenario, which could reduce the size of the uncertainties encountered in the analysis of the reference discharge, in particular when employing higher modulation frequencies.

Discharge \#42339 was designed to reproduce the pedestal-top density values of \#29216 and is referred to in the following as the high-$n_e$, high-ECRH case ($5.09-7.09$~s). In contrast, discharge \#42340 was performed with significantly reduced gas fueling and is referred to as the low-$n_e$, high-ECRH case ($6.09-7.89$~s).

\begin{figure}[t]
    \centering
    \includegraphics[width=0.9\textwidth]{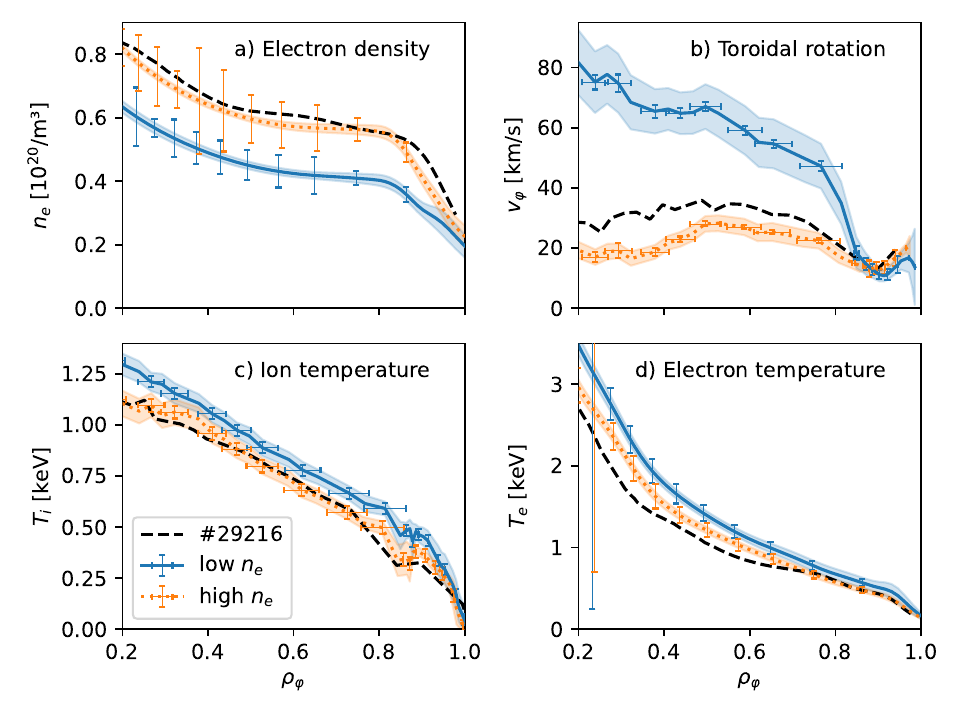}
    \caption{Kinetic profiles for the high-$n_e$ (\#42339, $5.09$--$7.09$~s, dotted orange) and low-$n_e$ (\#42340, $6.09$--$7.89$~s, solid blue) reproductions of the NBI+ECRH phase of \#29216 (black dashed). The high-$n_e$ case closely resembles \#29216 and reproduces the hollow rotation profile, while the low-$n_e$ case exhibits a more peaked rotation profile. The shaded bands indicate the standard deviation across the analyzed time windows at each radial location.}
    \label{fig:kinetic_profiles_new}
\end{figure}

Panel~(a) of Fig.~\ref{fig:kinetic_profiles_new} shows the electron density profiles for the high- and low-$n_e$ scenarios. The high-$n_e$ case closely matches the density profile of the NBI+ECRH phase of \#29216 in the core within error bars (see black dashed line), while the low-fueling case exhibits an overall downward shift of the core density and modified pedestal gradients. Despite these differences, the core density gradients are found to be similar, as discussed later. Panel~(b) presents the measured toroidal plasma rotation profiles. The high-$n_e$ case mimics the hollow rotation profile observed in \#29216. In contrast, the low-$n_e$ case exhibits comparable edge rotation values but increases toward significantly higher rotation levels in the core with only a small hollow ''dip" around $\rho_\varphi \approx 0.4$. This can be related to the fact that the same NBI torque density acts locally on fewer particles, which are then accelerated to higher rotation values. The precise rotation level, however, is set self-consistently by the interplay of all transport mechanisms and source terms.

As gradients appear to be very similar between the high- and low-$n_e$ case, the lower overall density must lead to changes in the relative size of the transport components, preventing the formation of hollow rotation profiles. Panels~(c) and~(d) show the ion and electron temperature profiles, respectively. While the gradients remain very similar, a slight upward shift is observed in the low-$n_e$ case, consistent with the idea of identical heating power distributed over fewer particles. A comparison of the temperature profiles with those of the reference discharge shows agreement within the error bars, although some gradients appear to differ slightly. A more precise reproduction of discharge \#29216 was not achieved. However, such exact agreement is not required to capture and understand the underlying dynamics associated with the density increase observed in these experiments, as discussed in the following.

\begin{figure}
    \centering
    \includegraphics[width=0.9\linewidth]{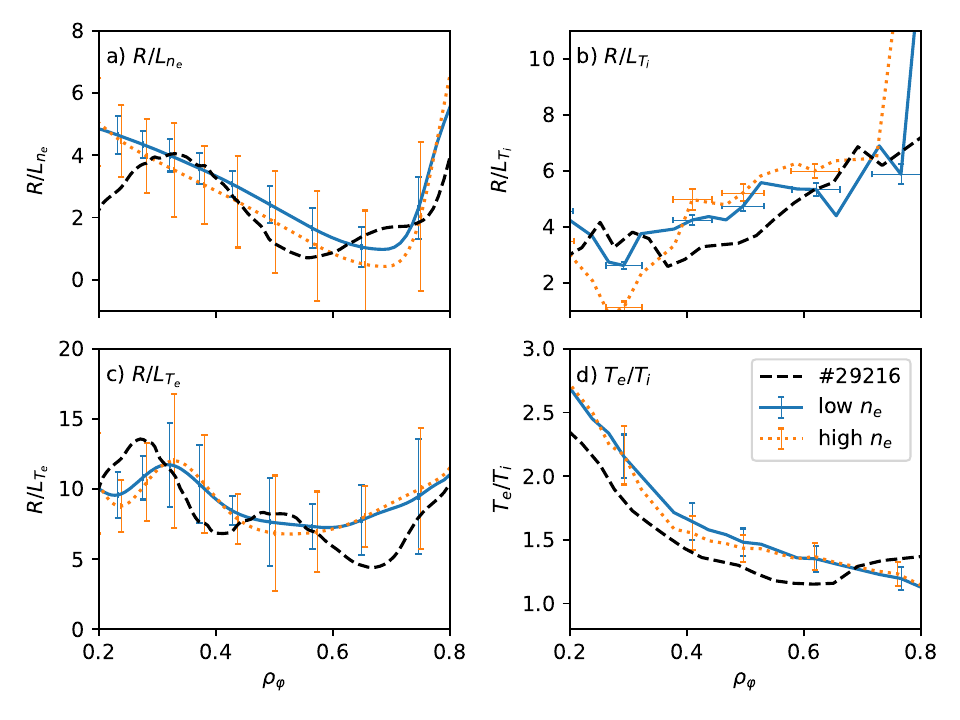}
    \caption{Selected dimensionless quantities for the high-$n_e$ (\#42339, $5.09$--$7.09$~s, dotted orange) and low-$n_e$ (\#42340, $6.09$--$7.89$~s, solid blue) reproductions of the NBI+ECRH phase of \#29216 (black dashed).}
    \label{fig:gradients_new}
\end{figure}

Next, Fig.~\ref{fig:gradients_new} confirms the previously stated observation that the gradients are very similar between the two discharge phases, with the exception of localized "dips" in the logarithmic ion temperature profiles, see Panel~(b). These "dips" are likely caused by disturbances in the CXRS spectra collected by the corresponding channel, combined with the limited radial smoothing applied for most accurate data analysis. Not shown are $q$, $\beta_e$, $\rho_*$, and $\nu_*$, which show moderate responses to the changed density but are otherwise very similar to those shown in Fig. \ref{fig:core_gradients}. Given the overall agreement within experimental uncertainties in the estimation of gradients, both discharges are expected to exhibit very similar turbulence characteristics. This motivates the assumption that they could indeed exhibit very similar Prandtl and pinch numbers, as well as comparable intrinsic torque levels. A comparison with the NBI+ECRH phase of \#29216 further shows general agreement, suggesting that similar momentum transport behavior can be expected in these new discharges.

\begin{figure}[t]
    \centering
    \includegraphics[width=0.9\textwidth]{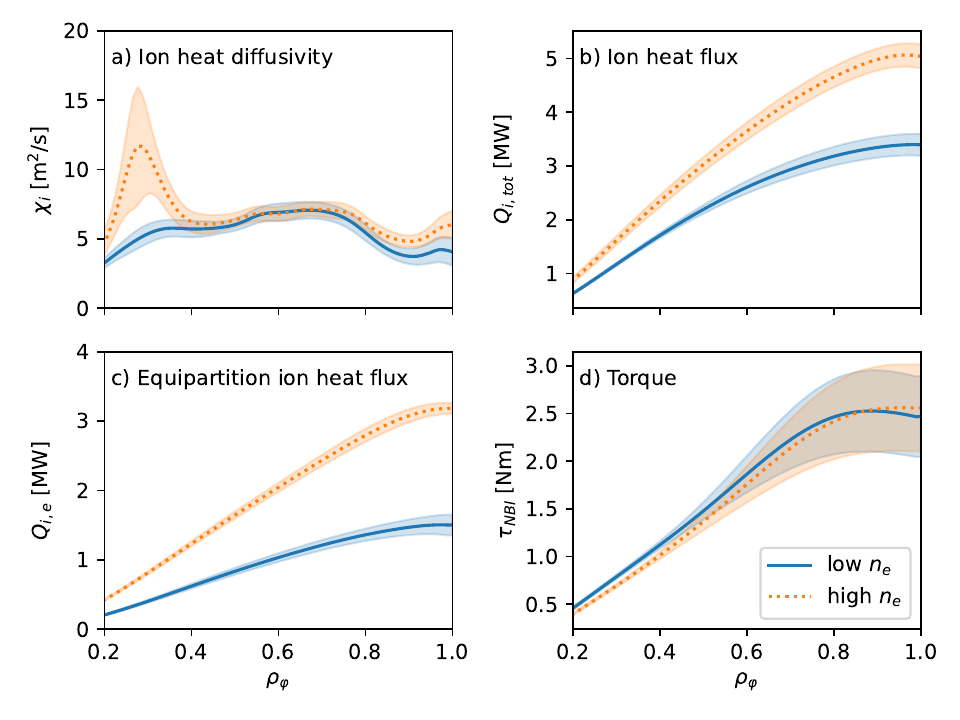}
    \caption{Ion heat diffusivity (a) from ASTRA calculations. Ion heat fluxes (b), equipartition heat fluxes (c), and applied torque (d) from TRANSP calculations, shown for the high $n_e$ (\#42339, $5.09-7.09$s, dotted orange) and low $n_e$ (\#42340, $6.09-7.89$s, solid blue) reproduction of \#29216. The band structures indicate the standard deviation across the analyzed time windows for each radial point.}
    \label{fig:fluxes_new}
\end{figure}

Figure~\ref{fig:fluxes_new} shows the corresponding heat and torque fluxes. Panel~(a) displays the ion heat diffusivity, which is very similar for both discharges. A small peak at smaller radii is observed, linked to local flattenings in $\nabla T_i$, as shown in Fig. \ref{fig:gradients_new}(b). In general, the ion heat flux shown in Panel~(b) is significantly larger in the high-$n_e$ case, and the equipartition heat flux in Panel~(c) indicates stronger electron-ion coupling at higher density, as expected. However, since the case with higher heat flux also has higher density, the resulting ion heat diffusivities from power balance are similar, as shown. Panel~(d) shows that the externally applied NBI torque is also very similar between the two discharges. In the low-$n_e$ case, the torque penetrates slightly deeper into the plasma core, but this effect seems modest.

When combining the nearly identical applied torque with the different density profiles shown in Fig.~\ref{fig:kinetic_profiles_new}(a), the pronounced differences in pedestal-top rotation values in Fig.~\ref{fig:kinetic_profiles_new}(b) can be explained by the diverging density profiles. The data therefore indicate that the density variation, with a corresponding change in inertia, is the primary cause of the differing pedestal-top rotation levels between the two discharges.

As similar dimensionless parameters are observed for all three discharges, see discussion of Fig. \ref{fig:gradients_new},it can be assumed that comparable transport dynamics are present in both the high- and low-$n_e$ reproductions of \#29216. This observation supports a scenario in which a strong inward convective momentum flux, such as that inferred for the NBI+ECRH phase of \#29216, see Fig.~\ref{fig:flux_comparison}(b), increases significantly with the background rotation level. In the low-$n_e$ case, the higher pedestal-top rotation amplifies the inward convective flux, enabling it to balance the counter-current intrinsic torque. Consequently, the formation of hollow rotation profiles is largely suppressed, in agreement with the experimentally observed non-hollow rotation during the low-$n_e$ NBI+ECRH phase.

\begin{figure}[t]
    \centering
        \begin{overpic}[width=0.9\linewidth]{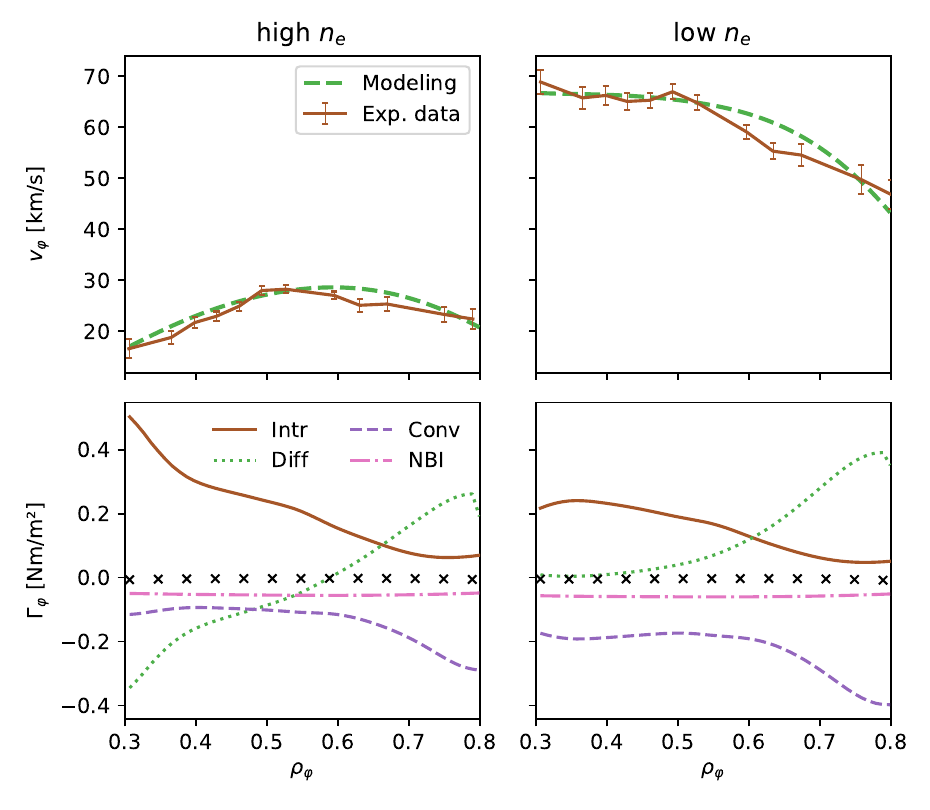}
        \put(97,10){\rotatebox{90}{\normalsize\sffamily ← inward}}
        \put(97,29){\rotatebox{90}{\normalsize\sffamily outward →}}
    \end{overpic}
    \caption{Application of the transport coefficients from the NBI+ECRH phase of \#29216 (as shown in Fig.~\ref{fig:transport_coefficients}) to the high-$n_e$ (\#42339, $5.09-7.09$~s, l.h.s.) and low-$n_e$ (\#42340, $6.09-7.89$~s, r.h.s.) scenarios. Despite the significant difference in the steady-state rotation profiles, both can be described with the same set of transport coefficients (see upper panels). The lower panels show the decomposition into the individual momentum flux contributions, with their sum indicated by black markers.}
    \label{fig:apply_29216_onto_new}
\end{figure}

This hypothesis can be tested by applying the transport coefficients fitted to the NBI+ECRH phase of \#29216, as shown in Fig.~\ref{fig:transport_coefficients}, to the corresponding discharge phases of \#42339 and \#42340. As shown in the upper Panels of Fig.~\ref{fig:apply_29216_onto_new}, and with respect to the involved error bars, this approach yields good agreement between the modeled and experimentally measured rotation profiles for both discharges over the radial range in which the original transport coefficients were fitted. The good agreement confirms that the dimensionless quantities are sufficiently similar to yield comparable transport dynamics across all three discharges. However, it should be noted that this does not imply a universal validity of the applied transport coefficients, but is rather a result of similar background profiles implying similar turbulence and transport characteristics.

A closer analysis of the resulting momentum fluxes is shown in the lower panels. The intrinsic fluxes (solid brown line) are similar, in the high-$n_e$ case, the intrinsic flux responds to the $\chi_i$ spike at the innermost radii (see Fig.~\ref{fig:fluxes_new}a), although this effect is rather localized. The same spike propagates into the diffusive flux (dotted green line), balancing the intrinsic component. Beyond this, the most significant difference is found in the convective momentum flux (dashed purple line), which increases over the entire radial range in the low-$n_e$ case and becomes the dominant contribution at larger radii. This suggests that the pedestal-top boundary condition, in combination with inward convective momentum transport, helps to prevent the formation of hollow rotation profiles, even in the presence of similar counter-current intrinsic torque. The black markers for both cases are very close to zero, confirming the self-consistency of the imposed combination of transport coefficients.

In conclusion, these additional discharges show that the emergence of hollow rotation profiles depends critically on the interplay between intrinsic torque and convective momentum transport. In scenarios with sufficiently high pedestal-top rotation, such as those arising from lower pedestal density at similar applied torque, the convective contribution can counteract the intrinsic torque and suppress the formation of hollow rotation profiles.

\section{Summary}
\label{sec:summary}

This work investigated toroidal momentum transport in type-I ELMy H-mode plasmas in the AUG tokamak, with a particular focus on understanding the formation of hollow rotation profiles under strong ECRH. Hollow rotation, characterized by a reversal of the rotation gradient in the plasma core, represents a critical scenario for future devices due to its increased susceptibility to locking and emergence of harmful MHD modes at resonant surfaces in the core. To address this issue, the established momentum transport analysis framework at AUG was applied to a NBI modulation experiment that features a plasma phase dominated by NBI heating and a second phase with strong additional ECRH at nearly unchanged externally applied background torque.

The momentum transport modeling self-consistently extracted diffusive, convective, and residual-stress contributions from steady-state, amplitude, and phase profiles of the modulated toroidal rotation. In both discharge phases, the experimentally measured profiles and time traces were reproduced with high accuracy, confirming the consistency of the inferred transport coefficients. The Prandtl number was found to be similar in both phases and to increase with radius, in agreement with previous AUG studies. Owing to the higher ion heat diffusivity during the NBI+ECRH phase, the inferred momentum diffusivity increased accordingly. Although the pinch number exhibited sizeable uncertainties, its overall trends remained consistent with an inward Coriolis momentum pinch correlated with the logarithmic density gradient. Most importantly, a strongly counter-current intrinsic torque was identified during the strong-ECRH phase, likely due to a change in the background turbulence and corresponding gradient profile shapes.

Linear gyrokinetic simulations using the GKW code supported this interpretation, revealing a transition from ITG-dominated turbulence in the NBI phase to an ITG-TEM mixed regime under strong ECRH. This transition is consistent with the experimentally observed inward particle pinch and with expectations for counter-current residual-stress generation driven by modified density and electron-temperature gradient profile shapes. A systematic scan of the modeling cost function in pinch/residual-stress parameter space demonstrated that solutions involving outward convection were consistently disfavored, while optimal solutions clustered around inward convection combined with counter-current intrinsic torque. Decomposition of the momentum flux further showed that, in the NBI+ECRH phase, the intrinsic torque dominates the momentum balance, accompanied by a substantial inward convective contribution, preventing an even stronger collapse or hollowing.

To further investigate the conditions required for the formation of hollow rotation profiles, two additional high-ECRH discharges were performed with modified density achieved through gas fueling. Despite exhibiting similar core gradients (and, thus, transport), only the high-density case reproduced hollow rotation profiles. In contrast, the low-density case maintained higher pedestal-top rotation, due to the same applied input torque at lower density, and, interestingly, a non-hollow, peaked rotation profile. By applying the fitting results from the reference discharge onto these additional discharges, it was shown that the emergence of hollow rotation depends critically on the balance between counter-current intrinsic torque and inward convective momentum transport.

These results provide important constraints for predicting rotation behavior in future low-torque tokamak scenarios. From the perspective of future reactor operation, peaked rotation profiles with strong rotational shear are generally favorable. In the absence of strong external torque sources, this suggests that avoiding an ITG–TEM mixed turbulence regime, which can generate strong counter-current intrinsic torque, may be beneficial. However, this is practically impossible to control, since the electron and ion heat fluxes in a future reactor are comparable in magnitude, automatically leading to a mixed-mode turbulence regime. Alternatively, in an ECRH dominated reactor scenario, strong coupling between electrons and ions would allow the ion heat flux to increase relative to the electron heat flux, pushing the turbulence back toward an ITG-dominated regime. However, the immediate implications of these insights, such as the mechanisms of intrinsic torque generation from profile-shearing effects, for integrated modeling of future devices are somewhat limited, due to the significant uncertainties associated with accurately predicting second-order profile gradients for future machines.

Therefore, given the limited control over the turbulence regime in a reactor environment, strong inward convective momentum transport appears advantageous, as it can mitigate or prevent the formation of hollow rotation profiles. As demonstrated in this analysis, the effectiveness of inward convection depends critically on the overall rotation level set at the pedestal top. This motivates further investigation of mechanisms capable of generating or breaking edge intrinsic torque. For understanding rotation levels in future low-torque devices, it is therefore important to understand edge intrinsic torque generation, such as neoclassical toroidal viscosity driven by externally applied 3D magnetic perturbations \cite{Zhu_2006,Garofalo_2008,COLE_2011,Shaing_NF_2015}, mechanisms of preferential ion losses \cite{Stoltzfus-Dueck_PRL_2012}, or the effect of strong $E \times B$ shearing \cite{Staebler_PRL_2013,Casson_PoP_2009}.

\section*{Acknowledgments}
This work has been carried out within the framework of the EUROfusion Consortium, partially funded by the European Union via the Euratom Research and Training Programme (Grant Agreement No 101052200 -- EUROfusion). The Swiss contribution to this work has been funded by the Swiss State Secretariat for Education, Research and Innovation (SERI). Views and opinions expressed are however those of the author(s) only and do not necessarily reflect those of the European Union, the European Commission or SERI. Neither the European Union nor the European Commission nor SERI can be held responsible for them. Large Language Models were used during the preparation of this manuscript to assist with grammar checking and to improve readability. All scientific content, interpretations, and conclusions have been carefully verified by the authors.

\printbibliography

@article{buttery2008,
	title        = {Multimachine extrapolation of neoclassical tearing mode physics to ITER},
	author       = {Buttery, R. J. and Gerhardt, S and Isayama, A and La Haye, RJ and Strait, EJ and Brennan, DP and Buratti, P and Chandra, D and Coda, S and De Grassie, J and others},
	year         = 2008,
	journal      = {Proceedings of 22nd IAEA Fusion Energy Conference},
	url          = {https://infoscience.epfl.ch/record/135075/files/buttery_iaea_08.pdf}
}

@unpublished{Tardini2026ASTRA8,
  author       = {G. Tardini and others},
  title        = {ASTRA-8: a modern framework for transport analysis and modelling in fusion devices},
  note         = {Submitted to Plasma Physics and Controlled Fusion},
  year         = {2026}
}

@article{angioni2011gyrokinetic,
  title={Gyrokinetic modelling of electron and boron density profiles of H-mode plasmas in ASDEX Upgrade},
  author={Angioni, C and McDermott, RM and Fable, E and Fischer, R and P{\"u}tterich, T and Ryter, F and Tardini, G and ASDEX Upgrade Team},
  journal={Nucl. Fusion},
  volume={51},
  number={2},
  pages={023006},
  year={2011}
}

@article{Ida_2021_LHD,
doi = {10.1088/1741-4326/abbf62},
url = {https://doi.org/10.1088/1741-4326/abbf62},
year = {2020},
publisher = {IOP Publishing},
volume = {61},
number = {1},
pages = {016012},
author = {Ida, K. and Yoshinuma, M. and Tanaka, K. and Nakata, M. and Kobayashi, T. and Fujiwara, Y. and Sakamoto, R. and Motojima, G. and Masuzaki, S. and the LHD Experiment Group},
title = {Characteristics of plasma parameters and turbulence in the isotope-mixing and the non-mixing states in hydrogen–deuterium mixture plasmas in the large helical device},
journal = {Nucl. Fusion}
}

@article{Ida2013ReversalTorque,
  author    = {K. Ida and H. Lee and K. Nagaoka and M. Osakabe and C. Suzuki and M. Yoshinuma and R. Seki and M. Yokoyama and T. Akiyama},
  title     = {Reversal of intrinsic torque associated with the formation of an internal transport barrier},
  journal   = {Phys. Rev. Lett.},
  volume    = {111},
  number    = {5},
  pages     = {055001},
  year      = {2013},
  publisher = {American Physical Society},
  doi       = {10.1103/PhysRevLett.111.055001}
}

@article{Stoltzfus-Dueck_2012,
	title        = {Transport-Driven Toroidal Rotation in the Tokamak Edge},
	author       = {Stoltzfus-Dueck, T.},
	year         = 2012,
	journal      = {Phys. Rev. Lett.},
	publisher    = {American Physical Society},
	volume       = 108,
	pages        = {065002},
	doi          = {10.1103/PhysRevLett.108.065002},
	url          = {https://link.aps.org/doi/10.1103/PhysRevLett.108.065002}          ,
	issue        = 6,
	numpages     = 5
}

@article{Waltz_1995_PoP,
	title        = {Advances in the simulation of toroidal gyro‐Landau fluid model turbulence},
	author       = {Waltz, R. E.  and Kerbel,G. D.  and Milovich,J.  and Hammett,G. W.},
	year         = 1995,
	journal      = {Phys. Plasmas},
	volume       = 2,
	number       = 6,
	pages        = {2408--2416},
	doi          = {10.1063/1.871264},
	url          = {https://doi.org/10.1063/1.871264}    
}

@article{tala2016_EPS,
	title        = {Multi-machine experiments to study the parametric dependences of momentum transport and intrinsic torque},
	author       = {Tala, T. and Chrystal, C. and McDermott, R. M. and Pehkonen, S-P and Salmi, A and Angioni, C and Barnes, M and Duval, B and Giroud, C and Grierson, B and others},
	year         = 2016,
	journal      = {Proceedings of 43rd EPS Conference on Plasma Physics},
	url          = {https://pure.mpg.de/rest/items/item_2353276/component/file_3319023/content}
}

@article{Zimmermann2022,
	title        = {Analysis and modelling of momentum transport based on {NBI} modulation experiments at {ASDEX} Upgrade},
	author       = {C. F. B. Zimmermann and R M McDermott and E Fable and C Angioni and B P Duval and R Dux and A Salmi and U Stroth and T Tala and G Tardini and T P\"{u}tterich},
	year         = 2022,
	journal      = {Plasma Phys. Control. Fusion},
	publisher    = {{IOP} Publishing},
	volume       = 64,
	number       = 5,
	pages        = {055020},
	doi          = {10.1088/1361-6587/ac5ae8},
	url          = {https://doi.org/10.1088/1361-6587/ac5ae8}  
}

@article{COLE_2011,
    author = {Cole, A. J. and Callen, J. D. and Solomon, W. M. and Garofalo, A. M. and Hegna, C. C. and Lanctot, M. J. and Reimerdes, H.},
    title = {Peak neoclassical toroidal viscosity at low toroidal rotation in the DIII-D tokamaka)},
    journal = {Physics of Plasmas},
    volume = {18},
    number = {5},
    pages = {055711},
    year = {2011},
    issn = {1070-664X},
    doi = {10.1063/1.3590933},
    url = {https://doi.org/10.1063/1.3590933},
    eprint = {https://pubs.aip.org/aip/pop/article-pdf/doi/10.1063/1.3590933/15957025/055711_1_online.pdf},
}

@article{Garofalo_2008,
  title = {Observation of Plasma Rotation Driven by Static Nonaxisymmetric Magnetic Fields in a Tokamak},
  author = {Garofalo, A. M. and Burrell, K. H. and DeBoo, J. C. and deGrassie, J. S. and Jackson, G. L. and Lanctot, M. and Reimerdes, H. and Schaffer, M. J. and Solomon, W. M. and Strait, E. J.},
  journal = {Phys. Rev. Lett.},
  volume = {101},
  issue = {19},
  pages = {195005},
  numpages = {4},
  year = {2008},
  publisher = {American Physical Society},
  doi = {10.1103/PhysRevLett.101.195005},
  url = {https://link.aps.org/doi/10.1103/PhysRevLett.101.195005}
}

@article{Zhu_2006,
  title = {Observation of Plasma Toroidal-Momentum Dissipation by Neoclassical Toroidal Viscosity},
  author = {Zhu, W. and Sabbagh, S. A. and Bell, R. E. and Bialek, J. M. and Bell, M. G. and LeBlanc, B. P. and Kaye, S. M. and Levinton, F. M. and Menard, J. E. and Shaing, K. C. and Sontag, A. C. and Yuh, H.},
  journal = {Phys. Rev. Lett.},
  volume = {96},
  issue = {22},
  pages = {225002},
  numpages = {4},
  year = {2006},
  publisher = {American Physical Society},
  doi = {10.1103/PhysRevLett.96.225002},
  url = {https://link.aps.org/doi/10.1103/PhysRevLett.96.225002}
}

@article{Shaing_NF_2015,
	title        = {Neoclassical plasma viscosity and transport processes in non-axisymmetric tori},
	author       = {K. C. Shaing and K. Ida and S.A. Sabbagh},
	year         = 2015,
	journal      = {Nucl. Fusion},
	publisher    = {IOP Publishing},
	volume       = 55,
	number       = 12,
	pages        = 125001,
	doi          = {10.1088/0029-5515/55/12/125001},
	url          = {https://dx.doi.org/10.1088/0029-5515/55/12/125001} 
}

@article{Camenen_PRL_2009,
	title        = {Transport of Parallel Momentum Induced by Current-Symmetry Breaking in Toroidal Plasmas},
	author       = {Camenen, Y. and Peeters, A. G. and Angioni, C. and Casson, F. J. and Hornsby, W. A. and Snodin, A. P. and Strintzi, D.},
	year         = 2009,
	journal      = {Phys. Rev. Lett.},
	publisher    = {American Physical Society},
	volume       = 102,
	pages        = 125001,
	doi          = {10.1103/PhysRevLett.102.125001},
	url          = {https://link.aps.org/doi/10.1103/PhysRevLett.102.125001} ,
	issue        = 12,
	numpages     = 4
}

@article{Casson_PoP_2009,
	title        = {{Anomalous parallel momentum transport due to E×B flow shear in a tokamak plasma}},
	author       = {Casson, F. J. and Peeters, A. G. and Camenen, Y. and Hornsby, W. A. and Snodin, A. P. and Strintzi, D. and Szepesi, G.},
	year         = 2009,
	journal      = {Phys. Plasmas},
	volume       = 16,
	number       = 9,
	pages        = {092303},
	doi          = {10.1063/1.3227650},
	issn         = {1070-664X},
	url          = {https://doi.org/10.1063/1.3227650} 
}

@article{Yoshida_2008_PRL,
	title        = {Role of Pressure Gradient on Intrinsic Toroidal Rotation in Tokamak Plasmas},
	author       = {Yoshida, M. and Kamada, Y. and Takenaga, H. and Sakamoto, Y. and Urano, H. and Oyama, N. and Matsunaga, G.},
	year         = 2008,
	journal      = {Phys. Rev. Lett.},
	publisher    = {American Physical Society},
	volume       = 100,
	pages        = 105002,
	doi          = {10.1103/PhysRevLett.100.105002},
	url          = {https://link.aps.org/doi/10.1103/PhysRevLett.100.105002}  ,
	collaboration = {the JT-60 Team},
	issue        = 10,
	numpages     = 4
}

@article{Petty1998,
	title        = {Experimental constraints on transport from dimensionless parameter scaling studies},
	author       = {Petty,C. C.  and Luce,T. C.  and Baker,D. R.  and Ballet,B.  and Carlstrom,T. N.  and Cordey,J. G.  and DeBoo,J. C.  and Gohil,P.  and Groebner,R. J.  and Rice,B. W.  and Thomas,D. M.  and Wade,M. R.  and Waltz,R. E.},
	year         = 1998,
	journal      = {Phys. Plasmas},
	volume       = 5,
	number       = 5,
	pages        = {1695--1702},
	doi          = {10.1063/1.872838},
	url          = {https://doi.org/10.1063/1.872838}  
}

@article{Angioni_2017,
	title        = {A comparison of the impact of central {ECRH} and central {ICRH} on the tungsten behaviour in {ASDEX} Upgrade H-mode plasmas},
	author       = {C. Angioni and M. Sertoli and R. Bilato and V. Bobkov and A. Loarte and R. Ochoukov and T. Odstrcil and T. Pütterich and J. Stober},
	year         = 2017,
	journal      = {Nucl. Fusion},
	publisher    = {{IOP} Publishing},
	volume       = 57,
	number       = 5,
	pages        = {056015},
	doi          = {10.1088/1741-4326/aa6453},
	url          = {https://doi.org/10.1088/1741-4326/aa6453}  
}

@article{Angioni_PRL_2011,
	title        = {Intrinsic Toroidal Rotation, Density Peaking, and Turbulence Regimes in the Core of Tokamak Plasmas},
	author       = {Angioni, C. and McDermott, R. M. and Casson, F. J. and Fable, E. and Bottino, A. and Dux, R. and Fischer, R. and Podoba, Y. and P\"utterich, T. and Ryter, F. and Viezzer, E.},
	year         = 2011,
	journal      = {Phys. Rev. Lett.},
	publisher    = {American Physical Society},
	volume       = 107,
	pages        = 215003,
	doi          = {10.1103/PhysRevLett.107.215003},
	url          = {https://link.aps.org/doi/10.1103/PhysRevLett.107.215003} ,
	collaboration = {ASDEX Upgrade Team},
	issue        = 21,
	numpages     = 5
}

@article{Degrassie2009PPCF,
	title        = {Tokamak rotation sources, transport and sinks},
	author       = {DeGrassie, J. S.},
	year         = 2009,
	journal      = {Plasma Phys. Control. Fusion},
	publisher    = {IOP Publishing},
	volume       = 51,
	number       = 12,
	pages        = 124047
}

@article{Lee2003,
	title        = {Observation of Anomalous Momentum Transport in Tokamak Plasmas with No Momentum Input},
	author       = {Lee, W. D. and Rice, J. E. and Marmar, E. S. and Greenwald, M. J. and Hutchinson, I. H. and Snipes, J. A.},
	year         = 2003,
	journal      = {Phys. Rev. Lett.},
	publisher    = {American Physical Society},
	volume       = 91,
	pages        = 205003,
	doi          = {10.1103/PhysRevLett.91.205003},
	url          = {https://link.aps.org/doi/10.1103/PhysRevLett.91.205003} ,
	issue        = 20,
	numpages     = 4
}

@article{Rice2004,
	title        = {Observations of anomalous momentum transport in Alcator C-Mod plasmas with no momentum input},
	author       = {J. E. Rice and W.D Lee and E.S Marmar and P.T Bonoli and R.S Granetz and M.J Greenwald and A.E Hubbard and I.H Hutchinson and J.H Irby and Y Lin and D Mossessian and J.A Snipes and S.M Wolfe and S.J Wukitch},
	year         = 2004,
	journal      = {Nucl. Fusion},
	publisher    = {{IOP} Publishing},
	volume       = 44,
	number       = 3,
	pages        = {379--386},
	doi          = {10.1088/0029-5515/44/3/001},
	url          = {https://doi.org/10.1088/0029-5515/44/3/001}   
}

@article{Tardini2009,
	title        = {Angular momentum studies with {NBI} modulation in {JET}},
	author       = {G. Tardini and J. Ferreira and P. Mantica and A.G. Peeters and T. Tala and K.D. Zastrow and M. Brix and C. Giroud and G.V. Pereverzev and},
	year         = 2009,
	journal      = {Nucl. Fusion},
	publisher    = {{IOP} Publishing},
	volume       = 49,
	number       = 8,
	pages        = {085010},
	doi          = {10.1088/0029-5515/49/8/085010},
	url          = {https://doi.org/10.1088/0029-5515/49/8/085010}   
}

@article{biglari1990influence,
	title        = {Influence of sheared poloidal rotation on edge turbulence},
	author       = {H. Biglari and P. H. Diamond and P. W. Terry},
	year         = 1990,
	journal      = {Physics of Fluids B: Plasma Physics},
	publisher    = {{AIP} Publishing},
	volume       = 2,
	number       = 1,
	pages        = {1--4},
	doi          = {10.1063/1.859529},
	url          = {https://doi.org/10.1063/1.859529}   
}

@article{Viezzer2012,
	title        = {High-resolution charge exchange measurements at {ASDEX} Upgrade},
	author       = {E. Viezzer and T. P\"{u}tterich and R. Dux and R. M. McDermott and},
	year         = 2012,
	journal      = {Review of Scientific Instruments},
	publisher    = {{AIP} Publishing},
	volume       = 83,
	number       = 10,
	pages        = 103501,
	doi          = {10.1063/1.4755810},
	url          = {https://doi.org/10.1063/1.4755810} 
}

@article{Politzer2008,
	title        = {Influence of toroidal rotation on transport and stability in hybrid scenario plasmas in {DIII}-D},
	author       = {P. A. Politzer and C.C. Petty and R.J. Jayakumar and T.C. Luce and M.R. Wade and J.C. DeBoo and J.R. Ferron and P. Gohil and C.T. Holcomb and A.W. Hyatt and J. Kinsey and R.J. La Haye and M.A. Makowski and T.W. Petrie},
	year         = 2008,
	journal      = {Nucl. Fusion},
	publisher    = {{IOP} Publishing},
	volume       = 48,
	number       = 7,
	pages        = {075001},
	doi          = {10.1088/0029-5515/48/7/075001},
	url          = {https://doi.org/10.1088/0029-5515/48/7/075001} 
}

@article{Reimerdes2007,
	title        = {Reduced Critical Rotation for Resistive-Wall Mode Stabilization in a Near-Axisymmetric Configuration},
	author       = {Reimerdes, H. and Garofalo, A. M. and Jackson, G. L. and Okabayashi, M. and Strait, E. J. and Chu, M. S. and In, Y. and La Haye, R. J. and Lanctot, M. J. and Liu, Y. Q. and Navratil, G. A. and Solomon, W. M. and Takahashi, H. and Groebner, R. J.},
	year         = 2007,
	journal      = {Phys. Rev. Lett.},
	publisher    = {American Physical Society},
	volume       = 98,
	pages        = {055001},
	doi          = {10.1103/PhysRevLett.98.055001},
	url          = {https://link.aps.org/doi/10.1103/PhysRevLett.98.055001} ,
	issue        = 5,
	numpages     = 4
}

@article{Garofalo2002,
	title        = {Sustained Stabilization of the Resistive-Wall Mode by Plasma Rotation in the DIII-D Tokamak},
	author       = {Garofalo, A. M. and Strait, E. J. and Johnson, L. C. and La Haye, R. J. and Lazarus, E. A. and Navratil, G. A. and Okabayashi, M. and Scoville, J. T. and Taylor, T. S. and Turnbull, A. D.},
	year         = 2002,
	journal      = {Phys. Rev. Lett.},
	publisher    = {American Physical Society},
	volume       = 89,
	pages        = 235001,
	doi          = {10.1103/PhysRevLett.89.235001},
	url          = {https://link.aps.org/doi/10.1103/PhysRevLett.89.235001} ,
	issue        = 23,
	numpages     = 4
}

@article{Strait1995,
	title        = {Wall Stabilization of High Beta Tokamak Discharges in DIII-D},
	author       = {Strait, E. J. and Taylor, T. S. and Turnbull, A. D. and Ferron, J. R. and Lao, L. L. and Rice, B. and Sauter, O.},
	year         = 1995,
	journal      = {Phys. Rev. Lett.},
	publisher    = {American Physical Society},
	volume       = 74,
	pages        = {2483--2486},
	doi          = {10.1103/PhysRevLett.74.2483},
	url          = {https://link.aps.org/doi/10.1103/PhysRevLett.74.2483}  ,
	issue        = 13,
	numpages     = {0}
}

@article{Tala_2009,
	title        = {Evidence of Inward Toroidal Momentum Convection in the JET Tokamak},
	author       = {Tala, T. and Zastrow, K.-D. and Ferreira, J. and Mantica, P. and Naulin, V. and Peeters, A. G. and Tardini, G. and Brix, M. and Corrigan, G. and Giroud, C. and Strintzi, D.},
	year         = 2009,
	journal      = {Phys. Rev. Lett.},
	publisher    = {American Physical Society},
	volume       = 102,
	pages        = {075001},
	doi          = {10.1103/PhysRevLett.102.075001},
	url          = {https://link.aps.org/doi/10.1103/PhysRevLett.102.075001}   ,
	issue        = 7,
	numpages     = 4
}

@article{Terry2000,
	title        = {Suppression of turbulence and transport by sheared flow},
	author       = {Terry, P. W.},
	year         = 2000,
	journal      = {Reviews of Modern Physics},
	publisher    = {American Physical Society},
	volume       = 72,
	pages        = {109--165},
	doi          = {10.1103/RevModPhys.72.109},
	url          = {https://link.aps.org/doi/10.1103/RevModPhys.72.109}   ,
	issue        = 1,
	numpages     = {0}
}

@article{Burrell1997,
	title        = {Effects of E{\texttimes}B velocity shear and magnetic shear on turbulence and transport in magnetic confinement devices},
	author       = {K. H. Burrell},
	year         = 1997,
	journal      = {Phys. Plasmas},
	publisher    = {{AIP} Publishing},
	volume       = 4,
	number       = 5,
	pages        = {1499--1518},
	doi          = {10.1063/1.872367},
	url          = {https://doi.org/10.1063/1.872367}   
}

@article{hinton1976theory,
	title        = {Theory of plasma transport in toroidal confinement systems},
	author       = {Hinton, F. L. and Hazeltine, R. D.},
	year         = 1976,
	journal      = {Reviews of Modern Physics},
	publisher    = {American Physical Society},
	volume       = 48,
	pages        = {239--308},
	doi          = {10.1103/RevModPhys.48.239},
	url          = {https://link.aps.org/doi/10.1103/RevModPhys.48.239} ,
	issue        = 2,
	numpages     = {0}
}

@article{Solomon2009,
	title        = {Advances in understanding the generation and evolution of the toroidal rotation profile on {DIII}-D},
	author       = {W. M. Solomon and K.H. Burrell and A.M. Garofalo and A.J. Cole and R.V. Budny and J.S. DeGrassie and W.W. Heidbrink and G.L. Jackson and M.J. Lanctot and R. Nazikian and H. Reimerdes and E.J. Strait and M.A. Van Zeeland and},
	year         = 2009,
	journal      = {Nucl. Fusion},
	publisher    = {{IOP} Publishing},
	volume       = 49,
	number       = 8,
	pages        = {085005},
	doi          = {10.1088/0029-5515/49/8/085005},
	url          = {https://doi.org/10.1088/0029-5515/49/8/085005} 
}

@article{Solomon2011,
	title        = {Characterization of intrinsic rotation drive on {DIII}-D},
	author       = {W. M. Solomon and K.H. Burrell and J.S. DeGrassie and J.A. Boedo and A.M. Garofalo and R.A. Moyer and S.H. Muller and C.C. Petty and H. Reimerdes},
	year         = 2011,
	journal      = {Nucl. Fusion},
	publisher    = {{IOP} Publishing},
	volume       = 51,
	number       = 7,
	pages        = {073010},
	doi          = {10.1088/0029-5515/51/7/073010},
	url          = {https://doi.org/10.1088/0029-5515/51/7/073010} 
}

@incollection{hawryluk1981empirical,
  title={An empirical approach to tokamak transport},
  author={Hawryluk, R. J.},
  booktitle={Phys. Plasmas close to thermonuclear conditions},
  pages={19--46},
  year={1981},
  publisher={Elsevier}
}

@article{Yoshida_2012_intermachine,
	title        = {Momentum transport studies from multi-machine comparisons},
	author       = {M. Yoshida and S. Kaye and J. Rice and W. Solomon and T. Tala and R.E. Bell and K.H. Burrell and J. Ferreira and Y. Kamada and D. McDonald and P. Mantica and Y. Podpaly and M.L. Reinke and Y. Sakamoto and A. Salmi and the ITPA Transport & Confinement Topical Group},
	year         = 2012,
	journal      = {Nucl. Fusion},
	publisher    = {IOP Publishing and International Atomic Energy Agency},
	volume       = 52,
	number       = 12,
	pages        = 123005,
	doi          = {10.1088/0029-5515/52/12/123005},
	url          = {https://dx.doi.org/10.1088/0029-5515/52/12/123005} 
}

@article{mcdermott2011core,
	title        = {Core momentum and particle transport studies in the ASDEX Upgrade tokamak},
	author       = {R. M. McDermott and C Angioni and R Dux and E Fable and T Pütterich and F Ryter and A Salmi and T Tala and G Tardini and E Viezzer and the ASDEX Upgrade Team},
	year         = 2011,
	journal      = {Plasma Phys. Control. Fusion},
	publisher    = {},
	volume       = 53,
	number       = 12,
	pages        = 124013,
	doi          = {10.1088/0741-3335/53/12/124013},
	url          = {https://dx.doi.org/10.1088/0741-3335/53/12/124013} 
}

@article{Tala_2011,
	title        = {Parametric dependencies of momentum pinch and {Prandtl} number in {JET}},
	author       = {T. Tala and A. Salmi and C. Angioni and F.J. Casson and G. Corrigan and J. Ferreira and C. Giroud and P. Mantica and V. Naulin and A.G. Peeters and W.M. Solomon and D. Strintzi and M. Tsalas and T.W. Versloot and P.C. de Vries and K. Zastrow},
	year         = 2011,
	journal      = {Nucl. Fusion},
	publisher    = {{IOP} Publishing},
	volume       = 51,
	number       = 12,
	pages        = 123002,
	doi          = {10.1088/0029-5515/51/12/123002},
	url          = {https://doi.org/10.1088%2F0029-5515%2F51%2F12%2F123002}
}

@article{McDermott_2014,
	title        = {Core intrinsic rotation behaviour in {ASDEX} {Upgrade} {Ohmic} {L}-mode plasmas},
	author       = {R. M. McDermott and C. Angioni and G.D. Conway and R. Dux and E. Fable and R. Fischer and T. Pütterich and F. Ryter and E. Viezzer},
	year         = 2014,
	journal      = {Nucl. Fusion},
	publisher    = {{IOP} Publishing},
	volume       = 54,
	number       = 4,
	pages        = {043009},
	doi          = {10.1088/0029-5515/54/4/043009},
	url          = {https://doi.org/10.1088%2F0029-5515%2F54%2F4%2F043009}
}

@article{Angioni2015,
	title        = {The impact of poloidal asymmetries on tungsten transport in the core of {JET} H-mode plasmas},
	author       = {C. Angioni and F. J. Casson and P. Mantica and T. P\"{u}tterich and M. Valisa and E. A. Belli and R. Bilato and C. Giroud and P. Helander and},
	year         = 2015,
	journal      = {Phys. Plasmas},
	publisher    = {{AIP} Publishing},
	volume       = 22,
	number       = 5,
	pages        = {055902},
	doi          = {10.1063/1.4919036},
	url          = {https://doi.org/10.1063/1.4919036} 
}

@article{Mantica_2010,
	title        = {Perturbative studies of toroidal momentum transport using neutral beam injection modulation in the Joint European Torus: Experimental results, analysis methodology, and first principles modeling},
	author       = {Mantica, P.  and others},
	year         = 2010,
	journal      = {Phys. Plasmas},
	volume       = 17,
	number       = 9,
	pages        = {092505},
	doi          = {10.1063/1.3480640}
}

@article{Ida_1998,
	title        = {Experimental studies of the physical mechanism determining the radial electric field and its radial structure in a toroidal plasma},
	author       = {K. Ida},
	year         = 1998,
	journal      = {Plasma Phys. Control. Fusion},
	publisher    = {},
	volume       = 40,
	number       = 8,
	pages        = 1429,
	doi          = {10.1088/0741-3335/40/8/002},
	url          = {https://dx.doi.org/10.1088/0741-3335/40/8/002}  
}

@article{Zimmermann_2024,
    author = {Zimmermann, C. F. B. and Angioni, C. and McDermott, R. M. and Duval, B. P. and Dux, R. and Fable, E. and Salmi, A. and Stroth, U. and Tala, T. and Tardini, G. and P√ºtterich, T. and ASDEX Upgrade Team},

    title = {Experimental validation of momentum transport theory in the core of H-mode plasmas in the ASDEX Upgrade tokamak},

    journal = {Phys. Plasmas},

    volume = {31},

    number = {4},

    pages = {042306},

    year = {2024},

}

@article{mcdermott2011effect,
	title        = {Effect of electron cyclotron resonance heating (ECRH) on toroidal rotation in ASDEX Upgrade H-mode discharges},
	author       = {R. M. McDermott and C Angioni and R Dux and A Gude and T Pütterich and F Ryter and G Tardini and the ASDEX Upgrade Team},
	year         = 2011,
	journal      = {Plasma Phys. Control. Fusion},
	publisher    = {},
	volume       = 53,
	number       = 3,
	pages        = {035007},
	doi          = {10.1088/0741-3335/53/3/035007},
	url          = {https://dx.doi.org/10.1088/0741-3335/53/3/035007} 
}

@article{fable2015toroidal,
	title        = {A toroidal momentum transport equation for axisymmetric 1D transport codes},
	author       = {E. Fable},
	year         = 2015,
	journal      = {Plasma Phys. Control. Fusion},
	publisher    = {IOP Publishing},
	volume       = 57,
	number       = 4,
	pages        = {045007},
	doi          = {10.1088/0741-3335/57/4/045007},
	url          = {https://dx.doi.org/10.1088/0741-3335/57/4/045007} 
}

@article{mlynek2010design,
	title        = {{Design of a digital multiradian phase detector and its application in fusion plasma interferometry}},
	author       = {Mlynek, A. and Schramm, G. and Eixenberger, H. and Sips, G. and McCormick, K. and Zilker, M. and Behler, K. and Eheberg, J. and ASDEX Upgrade Team},
	year         = 2010,
	journal      = {Review of Scientific Instruments},
	volume       = 81,
	number       = 3,
	pages        = {033507},
	doi          = {10.1063/1.3340944},
	issn         = {0034-6748},
	url          = {https://doi.org/10.1063/1.3340944} 
}

@article{Garbet_2004,
	title        = {Physics of transport in tokamaks},
	author       = {X. Garbet and P. Mantica and C Angioni and E Asp and Y Baranov and C Bourdelle and R Budny and F Crisanti and G Cordey and L Garzotti and N Kirneva and D Hogeweij and T Hoang and F Imbeaux and E Joffrin and X Litaudon and A Manini and D C McDonald and H Nordman and V Parail and A Peeters and F Ryter and C Sozzi and M Valovic and T Tala and A Thyagaraja and I Voitsekhovitch and J Weiland and H Weisen and A Zabolotsky and the JET EFDA Contributors},
	year         = 2004,
	journal      = {Plasma Phys. Control. Fusion},
	publisher    = {{IOP} Publishing},
	volume       = 46,
	number       = {12B},
	pages        = {B557 - B574},
	doi          = {10.1088/0741-3335/46/12b/045},
	url          = {https://doi.org/10.1088%2F0741-3335%2F46%2F12b%2F045}
}

@article{pankin2004,
	title        = {The tokamak Monte Carlo fast ion module NUBEAM in the National Transport Code Collaboration library},
	author       = {A. Pankin and Douglas McCune and Robert Andre and Glenn Bateman and Arnold Kritz},
	year         = 2004,
	journal      = {Comput. Phys. Commun.},
	volume       = 159,
	number       = 3,
	pages        = {157--184},
	doi          = {https://doi.org/10.1016/j.cpc.2003.11.002}  ,
	issn         = {0010-4655},
	url          = {https://www.sciencedirect.com/science/article/pii/S0010465504001109}
}

@article{angioni2014tungsten,
	title        = {Tungsten transport in JET H-mode plasmas in hybrid scenario, experimental observations and modelling},
	author       = {C. Angioni and P. Mantica and T. Pütterich and M. Valisa and M. Baruzzo and E.A. Belli and P. Belo and F.J. Casson and C. Challis and P. Drewelow and C. Giroud and N. Hawkes and T.C. Hender and J. Hobirk and T. Koskela and L. Lauro Taroni and C.F. Maggi and J. Mlynar and T. Odstrcil and M.L. Reinke and M. Romanelli and JET EFDA Contributors},
	year         = 2014,
	journal      = {Nucl. Fusion},
	publisher    = {IOP Publishing},
	volume       = 54,
	number       = 8,
	pages        = {083028},
	doi          = {10.1088/0029-5515/54/8/083028},
	url          = {https://dx.doi.org/10.1088/0029-5515/54/8/083028}  
}

@article{casson2013,
	title        = {Validation of gyrokinetic modelling of light impurity transport including rotation in ASDEX Upgrade},
	author       = {F. J. Casson and R.M. McDermott and C. Angioni and Y. Camenen and R. Dux and E. Fable and R. Fischer and B. Geiger and P. Manas and L. Menchero and G. Tardini and the ASDEX Upgrade Team},
	year         = 2013,
	journal      = {Nucl. Fusion},
	publisher    = {IOP Publishing and International Atomic Energy Agency},
	volume       = 53,
	number       = 6,
	pages        = {063026},
	doi          = {10.1088/0029-5515/53/6/063026},
	url          = {https://dx.doi.org/10.1088/0029-5515/53/6/063026}  
}

@article{Peeters2004,
	title        = {The effect of a uniform radial electric field on the toroidal ion temperature gradient mode},
	author       = {Peeters, A. G.  and Strintzi,D.},
	year         = 2004,
	journal      = {Phys. Plasmas},
	volume       = 11,
	number       = 8,
	pages        = {3748--3751},
	doi          = {10.1063/1.1762876},
	url          = {https://doi.org/10.1063/1.1762876}  
}

@article{McDermott2017,
	title        = {Extensions to the charge exchange recombination spectroscopy diagnostic suite at {ASDEX} {Upgrade}},
	author       = {R. M. McDermott and A. Lebschy and B. Geiger and C. Bruhn and M. Cavedon and M. Dunne and R. Dux and R. Fischer and A. Kappatou and T. P\"{u}tterich and E. Viezzer},
	year         = 2017,
	journal      = {Review of Scientific Instruments},
	publisher    = {{AIP} Publishing},
	volume       = 88,
	number       = 7,
	doi          = {10.1063/1.4993131},
	url          = {https://doi.org/10.1063/1.4993131} 
}

@article{IDA_paper,
	title        = {Integrated Data Analysis of Profile Diagnostics at {ASDEX} {U}pgrade},
	author       = {R. Fischer and C. J. Fuchs and B. Kurzan and W. Suttrop and E. Wolfrum and {the ASDEX {U}pgrade Team}},
	year         = 2010,
	journal      = {Fusion Science and Technology},
	publisher    = {Taylor & Francis},
	volume       = 58,
	number       = 2,
	pages        = {675--684},
	doi          = {10.13182/FST10-110},
	url          = {https://doi.org/10.13182/FST10-110}    
}

@article{de_Vries_2011,
	title        = {Survey of disruption causes at {JET}},
	author       = {P. C. de Vries and M.F. Johnson and B. Alper and P. Buratti and T.C. Hender and H.R. Koslowski and V. Riccardo and},
	year         = 2011,
	journal      = {Nucl. Fusion},
	publisher    = {{IOP} Publishing},
	volume       = 51,
	number       = 5,
	pages        = {053018},
	doi          = {10.1088/0029-5515/51/5/053018},
	url          = {https://doi.org/10.1088/0029-5515/51/5/053018}    
}

@article{Diamond2009,
	title        = {Physics of non-diffusive turbulent transport of momentum and the origins of spontaneous rotation in tokamaks},
	author       = {P. H. Diamond and C.J. McDevitt and \"O.D. Gürcan and T.S. Hahm and W. X. Wang and E.S. Yoon and I. Holod and Z. Lin and V. Naulin and R. Singh},
	year         = 2009,
	journal      = {Nucl. Fusion},
	publisher    = {{IOP} Publishing},
	volume       = 49,
	number       = 4,
	pages        = {045002},
	doi          = {10.1088/0029-5515/49/4/045002},
	url          = {https://doi.org/10.1088%2F0029-5515%2F49%2F4%2F045002}
}

@article{Hornsby2018,
	title        = {Global gyrokinetic simulations of intrinsic rotation in {ASDEX} {Upgrade} {Ohmic L-mode plasmas}},
	author       = {W. A. Hornsby and C. Angioni and Z.X. Lu and E. Fable and I. Erofeev and R. M. McDermott and A. Medvedeva and A. Lebschy and A.G. Peeters},
	year         = 2018,
	journal      = {Nucl. Fusion},
	publisher    = {{IOP} Publishing},
	volume       = 58,
	number       = 5,
	pages        = {056008},
	doi          = {10.1088/1741-4326/aab22f},
	url          = {https://doi.org/10.1088%2F1741-4326%2Faab22f}
}

@article{Wang_PRL_2009,
	title        = {Gyrokinetic Studies on Turbulence-Driven and Neoclassical Nondiffusive Toroidal-Momentum Transport and the Effect of Residual Fluctuations in Strong $E\ifmmode\times\else\texttimes\fi{}B$ Shear},
	author       = {Wang, W. X. and Hahm, T. S. and Ethier, S. and Rewoldt, G. and Lee, W. W. and Tang, W. M. and Kaye, S. M. and Diamond, P. H.},
	year         = 2009,
	journal      = {Phys. Rev. Lett.},
	publisher    = {American Physical Society},
	volume       = 102,
	pages        = {035005},
	doi          = {10.1103/PhysRevLett.102.035005},
	url          = {https://link.aps.org/doi/10.1103/PhysRevLett.102.035005}   ,
	issue        = 3,
	numpages     = 4
}

@article{Grierson_PRL_2017,
	title        = {Main-Ion Intrinsic Toroidal Rotation Profile Driven by Residual Stress Torque from Ion Temperature Gradient Turbulence in the DIII-D Tokamak},
	author       = {Grierson, B. A. and Wang, W. X. and Ethier, S. and Staebler, G. M. and Battaglia, D. J. and Boedo, J. A. and deGrassie, J. S. and Solomon, W. M.},
	year         = 2017,
	journal      = {Phys. Rev. Lett.},
	publisher    = {American Physical Society},
	volume       = 118,
	pages        = {015002},
	doi          = {10.1103/PhysRevLett.118.015002},
	url          = {https://link.aps.org/doi/10.1103/PhysRevLett.118.015002}   ,
	issue        = 1,
	numpages     = 5
}

@article{Wang_PoP_2010,
	title        = {Nonlinear flow generation by electrostatic turbulence in tokamaks},
	author       = {Wang,W. X.  and Diamond,P. H.  and Hahm,T. S.  and Ethier,S.  and Rewoldt,G.  and Tang,W. M.},
	year         = 2010,
	journal      = {Phys. Plasmas},
	volume       = 17,
	number       = 7,
	pages        = {072511},
	doi          = {10.1063/1.3459096},
	url          = {https://doi.org/10.1063/1.3459096}   
}

@article{Camenen2011,
	title        = {Consequences of profile shearing on toroidal momentum transport},
	author       = {Y. Camenen and Y. Idomura and S. Jolliet and A.G. Peeters},
	year         = 2011,
	journal      = {Nucl. Fusion},
	publisher    = {{IOP} Publishing},
	volume       = 51,
	number       = 7,
	pages        = {073039},
	doi          = {10.1088/0029-5515/51/7/073039},
	url          = {https://doi.org/10.1088%2F0029-5515%2F51%2F7%2F073039}
}

@article{Strintzi2008,
	title        = {The toroidal momentum diffusivity in a tokamak plasma: A comparison of fluid and kinetic calculations},
	author       = {D. Strintzi and A. G. Peeters and J. Weiland},
	year         = 2008,
	journal      = {Phys. Plasmas},
	publisher    = {{AIP} Publishing},
	volume       = 15,
	number       = 4,
	pages        = {044502},
	doi          = {10.1063/1.2907370},
	url          = {https://doi.org/10.1063/1.2907370} 
}

@article{TRANSP_Reference_Paper,
	title        = {TRANSP},
	author       = {Breslau,  J. and Gorelenkova,  Marina and Poli,  Francesca and Sachdev,  Jai and Yuan,  Xingqiu},
	year         = 2018,
	journal      = {Princeton Plasma Physics Laboratory,  Princeton,  New Jersey, USA},
	doi          = {10.11578/DC.20180627.4},
	url          = {https://www.osti.gov/doecode/biblio/12542}
}

@article{ASTRA_Reference_Paper,
	title        = {{ASTRA} {Automated System for TRansport Analysis} in a Tokamak},
	author       = {Pereverzev, G. and Yushmanov, P. N.},
	year         = 2002,
	journal      = {MPI for Plasma Physics, Garching, Germany},
	address      = {Germany},
	url          = {https://pure.mpg.de/rest/items/item_2138238/component/file_2138237/content}
}

@article{Poli2018,
	title        = {{TORBEAM} 2.0, a paraxial beam tracing code for electron-cyclotron beams in fusion plasmas for extended physics applications},
	author       = {E. Poli and A. Bock and M. Lochbrunner and O. Maj and M. Reich and A. Snicker and A. Stegmeir and F. Volpe and N. Bertelli and R. Bilato and G.D. Conway and D. Farina and F. Felici and L. Figini and R. Fischer and C. Galperti and T. Happel and Y.R. Lin-Liu and N.B. Marushchenko and U. Mszanowski and F.M. Poli and J. Stober and E. Westerhof and R. Zille and A.G. Peeters and G.V. Pereverzev},
	year         = 2018,
	journal      = {Comput. Phys. Commun.},
	publisher    = {Elsevier {BV}},
	volume       = 225,
	pages        = {36--46},
	doi          = {10.1016/j.cpc.2017.12.018},
	url          = {https://doi.org/10.1016/j.cpc.2017.12.018} 
}

@article{Zimmermann_NF_Isotope,
	title        = {Comparison of momentum transport in matched hydrogen and deuterium H-mode plasmas in ASDEX Upgrade},
	author       = {C. F. B. Zimmermann and R.M. McDermott and C. Angioni and B.P. Duval and R. Dux and E. Fable and A. Salmi and U. Stroth and T. Tala and G. Tardini and T. Pütterich and the ASDEX Upgrade Team},
	year         = 2023,
	journal      = {Nucl. Fusion},
	publisher    = {IOP Publishing},
	volume       = 63,
	number       = 12,
	pages        = 126006,
	doi          = {10.1088/1741-4326/acf387},
	url          = {https://dx.doi.org/10.1088/1741-4326/acf387} 
}

@article{Zimmermann_NF_Letter,
doi = {10.1088/1741-4326/ad0489},
url = {https://dx.doi.org/10.1088/1741-4326/ad0489} ,
year = {2023},
publisher = {IOP Publishing},
volume = {63},
number = {12},
pages = {124003},
author = {C. F. B. Zimmermann and R.M. McDermott and C. Angioni and B.P. Duval and R. Dux and E. Fable and T. Luda and A. Salmi and U. Stroth and T. Tala and G. Tardini and T. Pütterich and the ASDEX Upgrade Team},
title = {Experimental determination of the three components of toroidal momentum transport in the core of a tokamak plasma},
journal = {Nucl. Fusion}
}

@article{Peeters2005,
	title        = {Linear gyrokinetic calculations of toroidal momentum transport in a tokamak due to the ion temperature gradient mode},
	author       = {A. G. Peeters and C. Angioni},
	year         = 2005,
	journal      = {Phys. Plasmas},
	publisher    = {{AIP} Publishing},
	volume       = 12,
	number       = 7,
	pages        = {072515},
	doi          = {10.1063/1.1949608},
	url          = {https://doi.org/10.1063/1.1949608} 
}

@article{Stober_2003,
	title        = {Dependence of particle transport on heating profiles in {ASDEX} Upgrade},
	author       = {J. Stober and R Dux and O Gruber and L Horton and P Lang and R Lorenzini and C Maggi and F Meo and R Neu and J.-M Noterdaeme and A Peeters and G Pereverzev and F Ryter and A.C.C Sips and A Stäbler and H Zohm and the ASDEX Upgrade Team},
	year         = 2003,
	journal      = {Nucl. Fusion},
	publisher    = {{IOP} Publishing},
	volume       = 43,
	number       = 10,
	pages        = {1265--1271},
	doi          = {10.1088/0029-5515/43/10/030},
	url          = {https://doi.org/10.1088/0029-5515/43/10/030} 
}

@article{Angioni_2012,
	title        = {Analytic formulae for centrifugal effects on turbulent transport of trace impurities in tokamak plasmas},
	author       = {Angioni,C.  and Casson,F. J.  and Veth,C.  and Peeters,A. G.},
	year         = 2012,
	journal      = {Phys. Plasmas},
	volume       = 19,
	number       = 12,
	pages        = 122311,
	doi          = {10.1063/1.4773051},
	url          = {https://doi.org/10.1063/1.4773051}  
}

@article{Thomson_System_AUG,
	title        = {The Thomson scattering systems of the ASDEX upgrade tokamak},
	author       = {Murmann,H.  and Götsch,S.  and Röhr,H.  and Salzmann,H.  and Steuer,K. H.},
	year         = 1992,
	journal      = {Review of Scientific Instruments},
	volume       = 63,
	number       = 10,
	pages        = {4941--4943},
	doi          = {10.1063/1.1143504},
	url          = {https://doi.org/10.1063/1.1143504}   
}

@article{Holod_2010,
	title        = {Effects of electron dynamics in toroidal momentum transport driven by ion temperature gradient turbulence},
	author       = {I. Holod and Z. Lin},
	year         = 2010,
	journal      = {Plasma Phys. Control. Fusion},
	publisher    = {},
	volume       = 52,
	number       = 3,
	pages        = {035002},
	doi          = {10.1088/0741-3335/52/3/035002},
	url          = {https://dx.doi.org/10.1088/0741-3335/52/3/035002}   
}

@article{Holod_2008,
	title        = {{Gyrokinetic particle simulations of toroidal momentum transport}},
	author       = {Holod, I. and Lin, Z.},
	year         = 2008,
	journal      = {Phys. Plasmas},
	volume       = 15,
	number       = 9,
	doi          = {10.1063/1.2977769},
	issn         = {1070-664X},
	url          = {https://doi.org/10.1063/1.2977769}   ,
	note         = {092302}
}

@article{weiland2009symmetry,
	title        = {Symmetry breaking effects of toroidicity on toroidal momentum transport},
	author       = {J. Weiland and R. Singh and H. Nordman and P. Kaw and A.G. Peeters and D. Strinzi},
	year         = 2009,
	journal      = {Nucl. Fusion},
	publisher    = {},
	volume       = 49,
	number       = 6,
	pages        = {065033},
	doi          = {10.1088/0029-5515/49/6/065033},
	url          = {https://dx.doi.org/10.1088/0029-5515/49/6/065033}   
}

@article{peeters2006toroidal,
	title        = {Toroidal momentum transport},
	author       = {A. G. Peeters and C Angioni and A Bottino and A Kallenbach and B Kurzan and C F Maggi and W Suttrop and the ASDEX Upgrade team},
	year         = 2006,
	journal      = {Plasma Phys. Control. Fusion},
	publisher    = {},
	volume       = 48,
	number       = {12B},
	pages        = {B413},
	doi          = {10.1088/0741-3335/48/12B/S39},
	url          = {https://dx.doi.org/10.1088/0741-3335/48/12B/S39}  
}

@article{Hahm_PoP_2008,
	title        = {{Turbulent equipartition theory of toroidal momentum pinch}},
	author       = {Hahm, T. S. and Diamond, P. H. and Gurcan, O. D. and Rewoldt, G.},
	year         = 2008,
	journal      = {Phys. Plasmas},
	volume       = 15,
	number       = 5,
	pages        = {055902},
	doi          = {10.1063/1.2839293},
	issn         = {1070-664X},
	url          = {https://doi.org/10.1063/1.2839293}  
}

@article{Hahm1995,
	title        = {Flow shear induced fluctuation suppression in finite aspect ratio shaped tokamak plasma},
	author       = {T. S. Hahm and K. H. Burrell},
	year         = 1995,
	journal      = {Phys. Plasmas},
	publisher    = {{AIP} Publishing},
	volume       = 2,
	number       = 5,
	pages        = {1648--1651},
	doi          = {10.1063/1.871313},
	url          = {https://doi.org/10.1063/1.871313}  
}

@article{Hahm1994,
	title        = {Rotation shear induced fluctuation decorrelation in a toroidal plasma},
	author       = {T. S. Hahm},
	year         = 1994,
	journal      = {Phys. Plasmas},
	publisher    = {{AIP} Publishing},
	volume       = 1,
	number       = 9,
	pages        = {2940--2944},
	doi          = {10.1063/1.870534},
	url          = {https://doi.org/10.1063/1.870534}  
}

@article{Hahm_PoP_2007,
	title        = {{Nonlinear gyrokinetic theory of toroidal momentum pinch}},
	author       = {Hahm, T. S. and Diamond, P. H. and Gurcan, O. D. and Rewoldt, G.},
	year         = 2007,
	journal      = {Phys. Plasmas},
	volume       = 14,
	number       = 7,
	pages        = {072302},
	doi          = {10.1063/1.2743642},
	issn         = {1070-664X},
	url          = {https://doi.org/10.1063/1.2743642}  
}

@article{waltz2007coupled,
	title        = {{Coupled ion temperature gradient and trapped electron mode to electron temperature gradient mode gyrokinetic simulationsa)}},
	author       = {Waltz, R. E. and Candy, J. and Fahey, M.},
	year         = 2007,
	journal      = {Phys. Plasmas},
	volume       = 14,
	number       = 5,
	pages        = {056116},
	doi          = {10.1063/1.2436851},
	issn         = {1070-664X},
	url          = {https://doi.org/10.1063/1.2436851}     
}

@article{Solomon_2010,
	title        = {Mechanisms for generating toroidal rotation in tokamaks without external momentum input},
	author       = {Solomon,W. M.  and Burrell,K. H.  and Garofalo,A. M.  and Kaye,S. M.  and Bell,R. E.  and Cole,A. J.  and DeGrassie,J. S.  and Diamond,P. H.  and Hahm,T. S.  and Jackson,G. L.  and Lanctot,M. J.  and Petty,C. C.  and Reimerdes,H.  and Sabbagh,S. A.  and Strait,E. J.  and Tala,T.  and Waltz,R. E.},
	year         = 2010,
	journal      = {Phys. Plasmas},
	volume       = 17,
	number       = 5,
	pages        = {056108},
	doi          = {10.1063/1.3328521},
	url          = {https://doi.org/10.1063/1.3328521}        
}

@article{Peeters2007_PRL,
	title        = {Toroidal Momentum Pinch Velocity due to the {Coriolis} Drift Effect on Small Scale Instabilities in a Toroidal Plasma},
	author       = {A. G. Peeters and C. Angioni and D. Strintzi},
	year         = 2007,
	journal      = {Phys. Rev. Lett.},
	publisher    = {American Physical Society ({APS})},
	volume       = 98,
	number       = 26,
	pages        = 265003,
	doi          = {10.1103/physrevlett.98.265003},
	url          = {https://doi.org/10.1103/physrevlett.98.265003}       
}

@article{Peeters2011,
	title        = {Overview of toroidal momentum transport},
	author       = {A. G. Peeters and C. Angioni and A. Bortolon and Y. Camenen and F.J. Casson and B. Duval and L. Fiederspiel and W.A. Hornsby and Y. Idomura and T. Hein and N. Kluy and P. Mantica and F.I. Parra and A.P. Snodin and G. Szepesi and D. Strintzi and T. Tala and G. Tardini and P. de Vries and J. Weiland},
	year         = 2011,
	journal      = {Nucl. Fusion},
	publisher    = {{IOP} Publishing},
	volume       = 51,
	number       = 9,
	pages        = {094027},
	doi          = {10.1088/0029-5515/51/9/094027},
	url          = {https://doi.org/10.1088/0029-5515/51/9/094027}  
}

@article{Fable2013,
	title        = {Novel free-boundary equilibrium and transport solver with theory-based models and its validation against ASDEX Upgrade current ramp scenarios},
	author       = {E. Fable and C Angioni and F J Casson and D Told and A A Ivanov and F Jenko and R M McDermott and S Yu Medvedev and G V Pereverzev and F Ryter and W Treutterer and E Viezzer and the ASDEX Upgrade Team},
	year         = 2013,
	journal      = {Plasma Phys. Control. Fusion},
	publisher    = {IOP Publishing},
	volume       = 55,
	number       = 12,
	pages        = 124028,
	doi          = {10.1088/0741-3335/55/12/124028},
	url          = {https://dx.doi.org/10.1088/0741-3335/55/12/124028}   
}

@article{Yoshida_2009_PRL,
	title        = {Rotation Drive and Momentum Transport with Electron Cyclotron Heating in Tokamak Plasmas},
	author       = {Yoshida, M. and Sakamoto, Y. and Takenaga, H. and Ide, S. and Oyama, N. and Kobayashi, T. and Kamada, Y.},
	year         = 2009,
	journal      = {Phys. Rev. Lett.},
	publisher    = {American Physical Society},
	volume       = 103,
	pages        = {065003},
	doi          = {10.1103/PhysRevLett.103.065003},
	url          = {https://link.aps.org/doi/10.1103/PhysRevLett.103.065003} ,
	collaboration = {the JT-60 Team},
	issue        = 6,
	numpages     = 4
}

@article{Staebler_PRL_2013,
	title        = {New Paradigm for Suppression of Gyrokinetic Turbulence by Velocity Shear},
	author       = {Staebler, G. M. and Waltz, R. E. and Candy, J. and Kinsey, J. E.},
	year         = 2013,
	journal      = {Phys. Rev. Lett.},
	publisher    = {American Physical Society},
	volume       = 110,
	pages        = {055003},
	doi          = {10.1103/PhysRevLett.110.055003},
	url          = {https://link.aps.org/doi/10.1103/PhysR   evLett.110.055003},
	issue        = 5,
	numpages     = 5
}

@book{helander2002collisional,
	title        = {Collisional Transport in Magnetized Plasmas},
	author       = {Helander, P. and Sigmar, D. J.},
	year         = 2002,
	publisher    = {Cambridge University Press},
	series       = {Cambridge Monographs on Plasma Physics},
	isbn         = 9780521807982   ,
	lccn         = 2001035677
}

@article{lebschy2017measurement,
	title        = {Measurement of the complete core plasma flow across the LOC--SOC transition at ASDEX Upgrade},
	author       = {Lebschy, A. and McDermott, RM and Angioni, C and Geiger, B and Prisiazhniuk, D and Cavedon, M and Conway, GD and Dux, R and Dunne, MG and Kappatou, A and others},
	year         = 2017,
	journal      = {Nucl. Fusion},
	publisher    = {IOP Publishing},
	volume       = 58,
	number       = 2,
	pages        = {026013}
}

@article{Kluy_2009,
	title        = {{Linear gyrokinetic calculations of toroidal momentum transport in the presence of trapped electron modes in tokamak plasmas}},
	author       = {Kluy, N. and Angioni, C. and Camenen, Y. and Peeters, A. G.},
	year         = 2009,
	journal      = {Phys. Plasmas},
	volume       = 16,
	number       = 12,
	pages        = 122302,
	doi          = {10.1063/1.3271411},
	issn         = {1070-664X},
	url          = {https://doi.org/10.1063/1.3271411} 
}

@article{Peeters2009,
	title        = {The nonlinear gyro-kinetic flux tube code {GKW}},
	author       = {A. G. Peeters and Y. Camenen and F.J. Casson and W.A. Hornsby and A.P. Snodin and D. Strintzi and G. Szepesi},
	year         = 2009,
	journal      = {Comput. Phys. Commun.},
	volume       = 180,
	number       = 12,
	pages        = {2650--2672},
	doi          = {https://doi.org/10.1016/j.cpc.2009.07.001} ,
	issn         = {0010-4655},
	url          = {http://www.sciencedirect.com/science/article/pii/S0010465509002112}
}

@article{Fischer_2020,
	title        = {Estimation and Uncertainties of Profiles and Equilibria for Fusion Modeling Codes},
	author       = {R. Fischer and L. Giannone and J. Illerhaus and P. J. McCarthy and R. M. McDermott and ASDEX Upgrade Team},
	year         = 2020,
	journal      = {Fusion Science and Technology},
	publisher    = {Taylor & Francis},
	volume       = 76,
	number       = 8,
	pages        = {879--893},
	doi          = {10.1080/15361055.2020.1820794},
	url          = {https://doi.org/10.1080/15361055.2020.1820794} 
}

@article{Rice_2016,
doi = {10.1088/0741-3335/58/8/083001},
url = {https://dx.doi.org/10.1088/0741-3335/58/8/083001}                                    ,
year = {2016},
publisher = {IOP Publishing},
volume = {58},
number = {8},
pages = {083001},
author = {J. E. Rice},
title = {Experimental observations of driven and intrinsic rotation in tokamak plasmas},
journal = {Plasma Phys. Control. Fusion}
}

@book{rice_2022,
  title={Driven rotation, self-generated flow, and momentum transport in tokamak plasmas},
  author={Rice, J. E.},
  volume={119},
  year={2022},
  journal={Springer Nature}
}

@article{Ida_1995,
  title = {Evidence for a Toroidal-Momentum-Transport Nondiffusive Term from the JFT-2M Tokamak},
  author = {Ida, K. and Miura, Y. and Matsuda, T. and Itoh, K. and Hidekuma, S. and Itoh, S.-I. and {the JFT-2M Group}},
  journal = {Phys. Rev. Lett.},
  volume = {74},
  issue = {11},
  pages = {1990--1993},
  numpages = {0},
  year = {1995},
  publisher = {American Physical Society},
  doi = {10.1103/PhysRevLett.74.1990},
  url = {https://link.aps.org/doi/10.1103/PhysRevLett.74.1990}                 
}

@article{Rice_1998,
doi = {10.1088/0029-5515/38/1/306},
url = {https://dx.doi.org/10.1088/0029-5515/38/1/306}             ,
year = {1998},
publisher = {},
volume = {38},
number = {1},
pages = {75},
author = {J. E. Rice and  M. Greenwald and  I. H. Hutchinson and  E. S. Marmar and  Y. Takase and  S. M. Wolfe and  F. Bombarda},
title = {Observations of central toroidal rotation 
in ICRF heated Alcator C-Mod plasmas},
journal = {Nucl. Fusion}
}

@article{Ida_2014,
doi = {10.1088/0029-5515/54/4/045001},
url = {https://dx.doi.org/10.1088/0029-5515/54/4/     045001}                       ,
year = {2014},
publisher = {IOP Publishing},
volume = {54},
number = {4},
pages = {045001},
author = {K. Ida and J. E. Rice},
title = {Rotation and momentum transport in tokamaks and helical systems},
journal = {Nucl. Fusion}
}

@article{Stoltzfus-Dueck_PRL_2012,
  title = {Transport-Driven Toroidal Rotation in the Tokamak Edge},
  author = {Stoltzfus-Dueck, T.},
  journal = {Phys. Rev. Lett.},
  volume = {108},
  issue = {6},
  pages = {065002},
  numpages = {5},
  year = {2012},
  publisher = {American Physical Society},
  doi = {10.1103/PhysRevLett.108.065002},
  url = {https://link.aps.org/doi/10.1103/PhysRevLett.108.065002}
}

@article{Casson_2015,
	title        = {Theoretical description of heavy impurity transport and its application to the modelling of tungsten in {JET} and {ASDEX} upgrade},
	author       = {F. J. Casson and C. Angioni and E. A. Belli and R. Bilato and P. Mantica and T. Odstrcil and T. Pütterich and M. Valisa and L. Garzotti and C. Giroud and J. Hobirk and C. F. Maggi and J. Mlynar and M. L. Reinke and {JET-EFDA Contributors} and the {ASDEX Upgrade Team}},
	year         = 2015,
	journal      = {Plasma Phys. Control. Fusion},
	publisher    = {{IOP} Publishing},
	volume       = 57,
	number       = 1,
	pages        = {014031},
	doi          = {10.1088/0741-3335/57/1/014031},
	url          = {https://doi.org/10.1088/0741-3335/57/1/014031}           
}

@article{Angioni_2005,
    author = {Angioni, C. and Peeters, A. G. and Ryter, F. and Jenko, F. and Conway, G. D. and Dannert, T. and Fahrbach, H. U. and Reich, M. and Suttrop, W. and Fattorini, L. and the {ASDEX Upgrade Team}},
    title = "{Relationship between density peaking, particle thermodiffusion, Ohmic confinement, and microinstabilities in ASDEX Upgrade L-mode plasmas}",
    journal = {Physics of Plasmas},
    volume = {12},
    number = {4},
    pages = {040701},
    year = {2005},
    issn = {1070-664X},
    doi = {10.1063/1.1867492},
    url = {https://doi.org/10.1063/1.1867492}     
}

@article{Fable_2008,
    author = {Fable, E. and Angioni, C. and Sauter, O.},
    title = {Parametric dependence of particle pinch coefficients for electron particle transport in linear gyrokinetic theory},
    journal = {AIP Conference Proceedings},
    volume = {1069},
    number = {1},
    pages = {64-75},
    year = {2008},
    issn = {0094-243X},
    doi = {10.1063/1.3033732},
    url = {https://doi.org/10.1063/1.3033732},
    eprint = {https://pubs.aip.org/aip/acp/article-pdf/1069/1/64/11918014/64_1_online.pdf},
}

@article{Fable_2010,
doi = {10.1088/0741-3335/52/1/015007},
url = {https://doi.org/10.1088/0741-3335/52/1/015007},
year = {2010},
publisher = {},
volume = {52},
number = {1},
pages = {015007},
author = {Fable, E and Angioni, C and Sauter, O},
title = {The role of ion and electron electrostatic turbulence in characterizing stationary particle transport in the core of tokamak plasmas},
journal = {Plasma Phys. Control. Fusion}
}

@article{LIB_paper_1,
	doi = {10.1088/0741-3335/50/8/085009},
	url = {https://doi.org/10.1088%2F0741-3335%2F50%2F8%2F085009},
	year = 2008,
	publisher = {{IOP} Publishing},
	volume = {50},
	pages = {085009},
	number = {8},
	author = {R. Fischer and E Wolfrum and J Schweinzer },
	title = {Probabilistic lithium beam data analysis},
	journal = {Plasma Phys. Control. Fusion},
}

@article{Bardoczi_2024,
doi = {10.1088/1741-4326/ad7787},
url = {https://doi.org/10.1088/1741-4326/ad7787},
year = {2024},
publisher = {IOP Publishing},
volume = {64},
number = {12},
pages = {126005},
author = {Bardoczi, L. and Richner, N.J. and Logan, N.C. and Strait, E.J. and Holcomb, C.T. and Zhu, J. and Rea, C.},
title = {The root cause of disruptive NTMs and paths to stable operation in DIII-D ITER baseline scenario plasmas},
journal = {Nucl. Fusion}
}

\end{document}